\documentclass[lettersize,journal]{IEEEtran}
\usepackage{amsmath,amsfonts}
\usepackage{algorithmic}
\usepackage{algorithm}
\usepackage{array}
\usepackage[caption=false,font=normalsize,labelfont=sf,textfont=sf]{subfig}
\usepackage{textcomp}
\usepackage{stfloats}
\usepackage{url}
\usepackage{verbatim}
\usepackage{graphicx}
\usepackage{cite}
\usepackage{multirow}
\usepackage{stfloats}
\usepackage{booktabs}
\usepackage{color}
\usepackage{hyperref}

\begin{document}

\title{Cross-modal Cognitive Consensus guided Audio-Visual Segmentation}

\author{Zhaofeng Shi, Qingbo Wu,~\IEEEmembership{Member,~IEEE}, Fanman Meng,~\IEEEmembership{Member,~IEEE}, Linfeng Xu,~\IEEEmembership{Member,~IEEE}, Hongliang Li,~\IEEEmembership{Senior Member,~IEEE}


\thanks{The authors are with the School of Information and Communication Engineering, University of Electronic Science and Technology of China, Chengdu 611731, China (email: zfshi@std.uestc.edu.cn; qbwu@uestc.edu.cn; fmmeng@uestc.edu.cn; lfxu@uestc.edu.cn; hlli@uestc.edu.cn). Corresponding authors: Qingbo Wu; Hongliang Li.}

}




\maketitle

\begin{abstract}
Audio-Visual Segmentation (AVS) aims to extract the sounding object from a video frame, which is represented by a pixel-wise segmentation mask for application scenarios such as multi-modal video editing, augmented reality, and intelligent robot systems. The pioneering work conducts this task through dense feature-level audio-visual interaction, which ignores the dimension gap between different modalities. More specifically, the audio clip could only provide a \textit{Global} semantic label in each sequence, but the video frame covers multiple semantic objects across different \textit{Local} regions, which leads to mislocalization of the representationally similar but semantically different object. In this paper, we propose a Cross-modal Cognitive Consensus guided Network (C3N) to align the audio-visual semantics from the global dimension and progressively inject them into the local regions via an attention mechanism. Firstly, a Cross-modal Cognitive Consensus Inference Module (C3IM) is developed to extract a unified-modal label by integrating audio/visual classification confidence and similarities of modality-agnostic label embeddings. Then, we feed the unified-modal label back to the visual backbone as the explicit semantic-level guidance via a Cognitive Consensus guided Attention Module (CCAM), which highlights the local features corresponding to the interested object. Extensive experiments on the Single Sound Source Segmentation (S4) setting and Multiple Sound Source Segmentation (MS3) setting of the AVSBench dataset demonstrate the effectiveness of the proposed method, which achieves state-of-the-art performance. Code is available at \href{https://github.com/ZhaofengSHI/AVS-C3N}{https://github.com/ZhaofengSHI/AVS-C3N}.

\end{abstract}


\section{Introduction}

Interesting object segmentation is fundamental for high-efficiency multimedia analysis. In recent years, object segmentation has been well explored for various visual signals, which strive to extract all objects or stuff from different granularity including the semantic segmentation \cite{long2015fully,gao2022fbsnet,yin2022contour}, instance segmentation \cite{hariharan2014simultaneous,yin2021bridging,li2021image}, and panoptic segmentation \cite{kirillov2019panoptic,li2021fully}. Meanwhile, these great efforts also bring us sweet trouble. That is, are all the objects/stuff interesting or necessary for the users in analyzing the prevalent multimedia data, which typically contains both audio and visual modalities? 

In recent years, many outstanding cross-modal learning methods \cite{li2022cross,radford2021learning,yang2022lavt,luddecke2022image,liu2023referring} have been developed to construct cross-modal consensus. In \cite{zhou2022audio}, Zhou \textit{et al.} made a new exploration toward Audio-Visual Segmentation (AVS), which focuses on extracting the sounding objects from a video frame based on their pixel-wise correspondence with multiple application scenarios such as extracting the sounding objects in videos and customizing the background for multi-modal video editing \cite{lee2022sound,lee2022sound2,fu2022m3l}, identifying and emphasizing sounding objects for augmented reality \cite{kaghat2020new,yang2022audio}, navigating towards the sounding object in the scene for intelligent robot systems \cite{wu2009surveillance,gan2020look,younes2023catch}, and so on. This new task provides multiple modalities to facilitate more specific interesting object extraction and a dense feature-level audio-visual interaction framework is proposed to achieve this target. Despite the success in integrating multi-modal features \cite{zhou2022audio,mao2023contrastive,hao2023improving,mao2023multimodal}, it is still challenging to establish the semantic correspondence between audio and visual modalities due to the dimension gap. More specifically, the audio clip could only provide a \textit{Global} semantic label in each sequence, but the video frame covers multiple semantic objects across different \textit{Local} regions. Such \textit{Global-Local} dimension gap is difficult to fill by the feature-level interaction and leads to incorrect localization due to representationally similar but semantically different objects (e.g. a normal van and an ambulance) in a frame. Some methods \cite{liu2023audiovisual,liu2023bavs} aim to utilize audio and visual semantics to filter the mask of the sounding object among the generated potential masks via a two-stage strategy, whereas the two-stage frameworks are inconvenient to deploy and the aligned semantics are unable to directly interact with representations of the visual branch to improve the generated mask proposals.

\begin{figure}[t]
\centering
\includegraphics[width=1.0\linewidth]{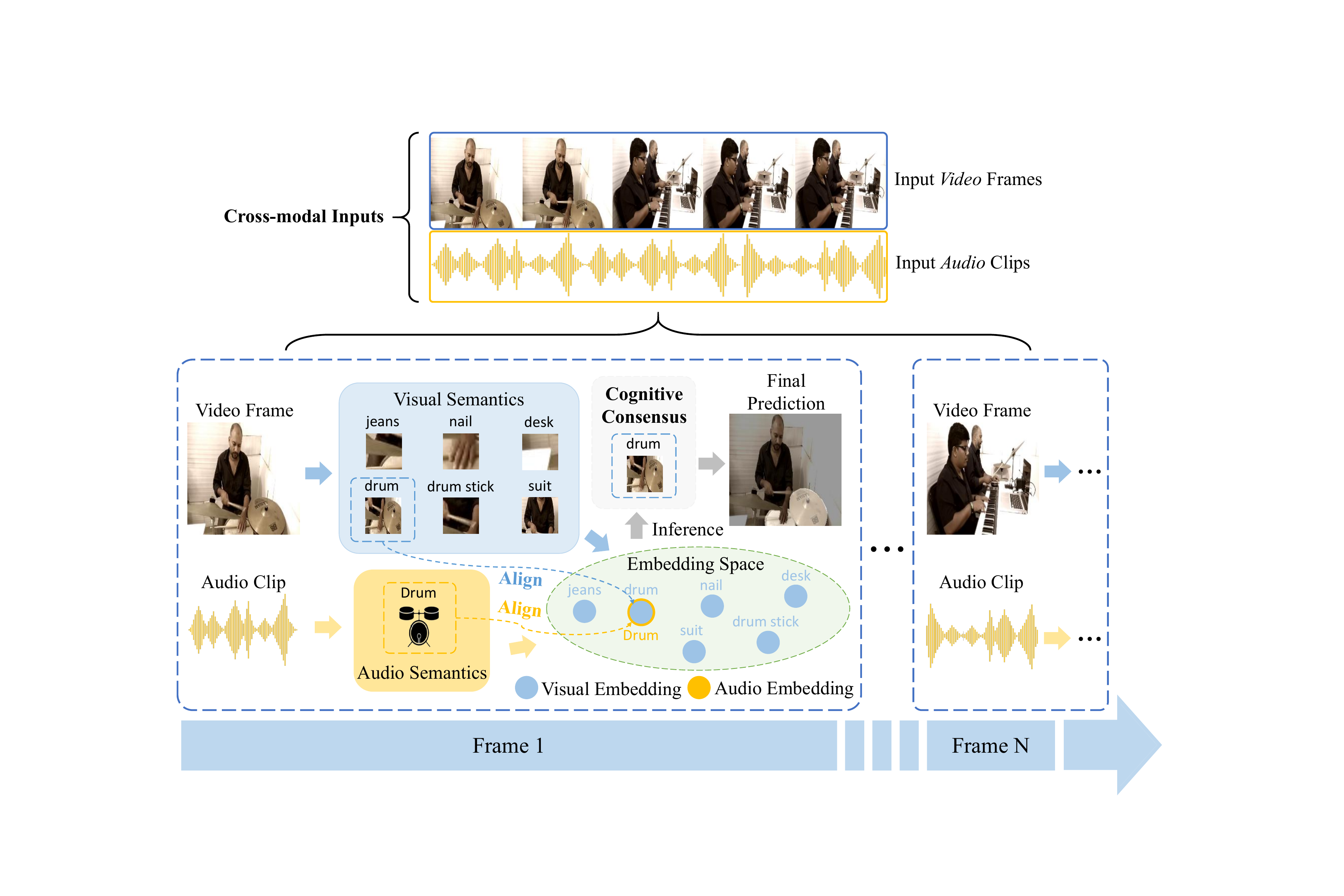}
\caption{Illustration of the proposed method. We first obtain the audio and visual semantics, which are then mapped into a unified embedding space. Based on the audio-visual semantic similarities, we infer the cognitive consensus as the guidance for the final segmentation.}
\label{fig:1}
\end{figure}

To overcome the aforementioned issues, we propose a novel Cross-modal Cognitive Consensus guided Network (C3N). Specifically, the ``cognitive consensus" refers to the unified-modal label, which conveys the same semantics across the inputs with different modalities. In the AVS field, the ``cognitive consensus" means the semantically consistent label of the audio and visual modalities, which is capable of two objectives including estimating the class label of the sounding object and highlighting its pixel-level spatial locations within the frame. By contrast, the traditional audio-visual synchronization or alignment methods \cite{chung2017out,lyu2023graph,chen2021audio,khosravan2019attention,wang2020alignnet} aim to match the non-uniform and irregular misaligned audio and visual events from the temporal positions and are unable to provide explicit audio-visual descriptive semantic information. The schematic diagram of our method is shown in Fig \ref{fig:1}. The C3N aligns audio and video from the global dimension via semantic-level information and injects the cognitive consensus into local visual regions through an attention mechanism for filling the ``dimension gap" and mitigating the mislocalization to improve the segmentation. Firstly, the proposed Cross-modal Cognitive Consensus Inference Module (C3IM) feeds the extracted visual and audio features into independent heads to obtain classification confidence and convert the modality-specific labels into embeddings to calculate the cross-modal semantic similarities. We multiply audio/visual confidence scores and semantic similarities to infer the cognitive consensus and extract a unified-modal label. Next, feed the unified-modal label back to hierarchical layers of the visual backbone as the global semantic-level guidance for highlighting local features corresponding to the object of interest in a channel-spatial attention manner through the parameter-free Cognitive Consensus guided Attention Module (CCAM). Finally, the refined features are fed into a cross-modal feature fusion module and a segmentation head for the segmentation.

The major contributions are concluded as follows:
\begin{itemize}
    \item We design a novel Cross-modal Cognitive Consensus Inference Module (C3IM) leveraging semantic-level audio/visual confidence and similarities of labels to infer cross-modal cognitive consensus and extract a unified-modal label for the subsequent segmentation guidance.
    \item The parameter-free Cognitive Consensus guided Attention Module (CCAM) is proposed to highlight the visual local feature elements corresponding to the sounding object with the guidance of the global unified-modal label in a channel-spatial attention manner.
    \item We propose a Cross-modal Cognitive Consensus guided Network (C3N) for the AVS task, whose key components are C3IM and CCAM. The results of the experiments show that C3N outperforms other methods and achieves state-of-the-art performance on the AVSBench dataset.
\end{itemize}

\section{Related Work}
\subsection{Visual Object Segmentation}
Researchers have proposed many methods for segmenting images or videos that only rely on visual input. Semantic Segmentation requires pixel-level category assignment of the given image, for which various CNN-based \cite{long2015fully,chen2017deeplab,zhao2017pyramid} and Transformer-based \cite{xie2021segformer,zheng2021rethinking,ru2022learning} models are well-designed with high accuracy. Moreover, Instance Segmentation \cite{hariharan2014simultaneous,he2017mask,liu2018path,zhang2022e2ec} aims to distinguish all objects into individual instances. Panoptic Segmentation \cite{kirillov2019panoptic,xiong2019upsnet,li2021fully} unifies the above two tasks, i.e. discriminating all objects and stuff in an image, and assigns semantic labels.

Video Object Segmentation (VOS) focuses on segmenting an object in a sequence of video frames. Semi-supervised VOS (SVOS) \cite{caelles2017one,oh2018fast,voigtlaender2019feelvos,yang2020collaborative,xu2020self,duke2021sstvos} means the annotation in the first frame is given as guidance during inference. OSVOS \cite{caelles2017one} relies on a pre-trained fully convolutional network. MaskTrack \cite{perazzi2017learning} and RGMP \cite{oh2018fast} are propagation-based methods, which make predictions based on the mask of the previous frame. VideoMatch \cite{hu2018videomatch} and A-GAME \cite{johnander2019generative} perform pixel-level match of frames to be segmented with the reference frame. FEELVOS \cite{voigtlaender2019feelvos} and CFBI \cite{yang2020collaborative} adopt the embedding learning method and perform pixel-level and instance-level matching between frames. SSTVOS \cite{duke2021sstvos} leverages the sparse attention mechanism to extract temporal-spatial relevance. Unsupervised VOS (UVOS) \cite{chen2015video,hu2018unsupervised,wang2019learning,lu2019see,ren2021reciprocal} means no human guidance information is available during model inference. Wang \textit{et al.} \cite{wang2019learning} and COSNet \cite{lu2019see} use attention-based methods to emphasize the correlations between video frames. Hu \textit{et al.} \cite{hu2018unsupervised}, Li \textit{et al.} \cite{li2018instance}, and Ren \textit{et al.} \cite{ren2021reciprocal} introduce additional motion cues during segmentation.

Despite the remarkable achievement of the visual object segmentation, the absence of information from other modalities prevents it from emphasizing objects of interest efficiently.

\subsection{Referring Video Object Segmentation}
Referring Video Object Segmentation (RVOS) aims to extract pixel-level segmentation masks of the interesting objects in the video according to the given language descriptions. Gavrilyuk \textit{et al.} \cite{gavrilyuk2018actor} make the first attempt towards the RVOS task. URVOS \cite{seo2020urvos} constructs a large-scale dataset and proposes a unified model, which utilizes the mask propagation operation. VTCapsule \cite{mcintosh2020visual} proposes a capsule-based method for learning effective representations. Some methods \cite{wang2019asymmetric,chen2022multi} integrate various attention mechanisms into RVOS frameworks to enhance the visual and text features. Ding \textit{et al.} \cite{ding2021progressive,ding2022language}, Feng \textit{et al.} \cite{feng2022deeply}, and Yang \textit{et al.} \cite{yang2021hierarchical} introduce hierarchical visual-language feature fusion methods. MTTR \cite{botach2022end} and ReferFormer \cite{wu2022language} model the task as a sequence prediction problem and propose Transformer-based networks to simplify the RVOS pipeline. Recently, Lan \textit{et al.} \cite{lan2024bidirectional} propose the awesome BIFIT framework, which performs inter-frame interaction and enhances the bilateral correlations between the linguistic and visual features. Luo \textit{et al.} \cite{luo2024soc} aggregate the semantic-level visual and textual information for temporal modeling and cross-modal alignment. Miao \textit{et al.} \cite{miao2023spectrum} use the spectrum-domain information for performing global interaction and extracting effective multimodal representations. 

Although the RVOS task introduces an additional text modality, it requires user interaction to provide linguistic guidance, whereas the AVS task leverages the inherent audio information from the input videos.

\subsection{Sound Source Localization and Segmentation}
Recently, researchers attempt to incorporate audio to extract the visual object of interest such as the Sound Source Localization (SSL) task aims to locate the sounding object. SSL evolves from Audio-Visual Correspondence (AVC) \cite{arandjelovic2017look}, which fuses audio and visual features to measure audio-visual consistency. With the growth of deep learning, many SSL methods \cite{senocak2018learning,arandjelovic2018objects,hu2019deep,qian2020multiple,chen2021localizing,song2022self,afouras2022self} have been proposed. Senocak \textit{et al.} \cite{senocak2018learning} adopts contrastive learning for audio-visual feature-level alignment. Chen \textit{et al.} \cite{chen2021localizing} and Lin \textit{et al.} \cite{lin2023unsupervised} improve localization performance through hard negative mining. Some recent methods \cite{hu2019deep,qian2020multiple,song2022self,afouras2022self,hu2020discriminative} try to localize in multi-source scenes. Hu \textit{et al.} \cite{hu2019deep,hu2020discriminative} propose clustering-based methods. Qian \textit{et al.} \cite{qian2020multiple} tackle this task by using the class-activated map (CAM). Song \textit{et al.} \cite{song2022self} propose a negative-free localization method achieved by mining explicit positive examples. Afouras \textit{et al.} \cite{afouras2020self}, Cheng \textit{et al.} \cite{cheng2020look} and Tian \textit{et al.} \cite{tian2021cyclic} propose general audio-visual frameworks, which are suitable for various downstream tasks including SSL. However, SSL generates heat maps for coarsely localizing sounding objects, which lack fine-grained descriptions.

Zhou \textit{et al.} \cite{zhou2022audio} first proposes the Audio-Visual Segmentation (AVS) task that predicts pixel-level masks for sounding objects, and an AVS baseline utilizes feature-level interactions. Since then, many ideas for AVS have been proposed, such as generation and reconstruction method \cite{hao2023improving}, latent diffusion-based method \cite{mao2023contrastive}, and representation learning \cite{mao2023multimodal}. However, due to the lack of semantic-level guidance, it remains hard to align audio and video explicitly. There are also some methods \cite{liu2023audiovisual,liu2023bavs} leverage semantic information in a two-stage strategy (i.e. first generate potential masks, then filter the mask of the sounding object using semantics) and BAVS\cite{liu2023bavs} uses the extra large model to extract semantics, while our C3N is an efficient end-to-end framework and the aligned semantics directly interact with representations of model middle layers to correct and improve the predict segmentation masks.

\begin{figure*}[!t]
\centering
\includegraphics[width=0.95\linewidth]{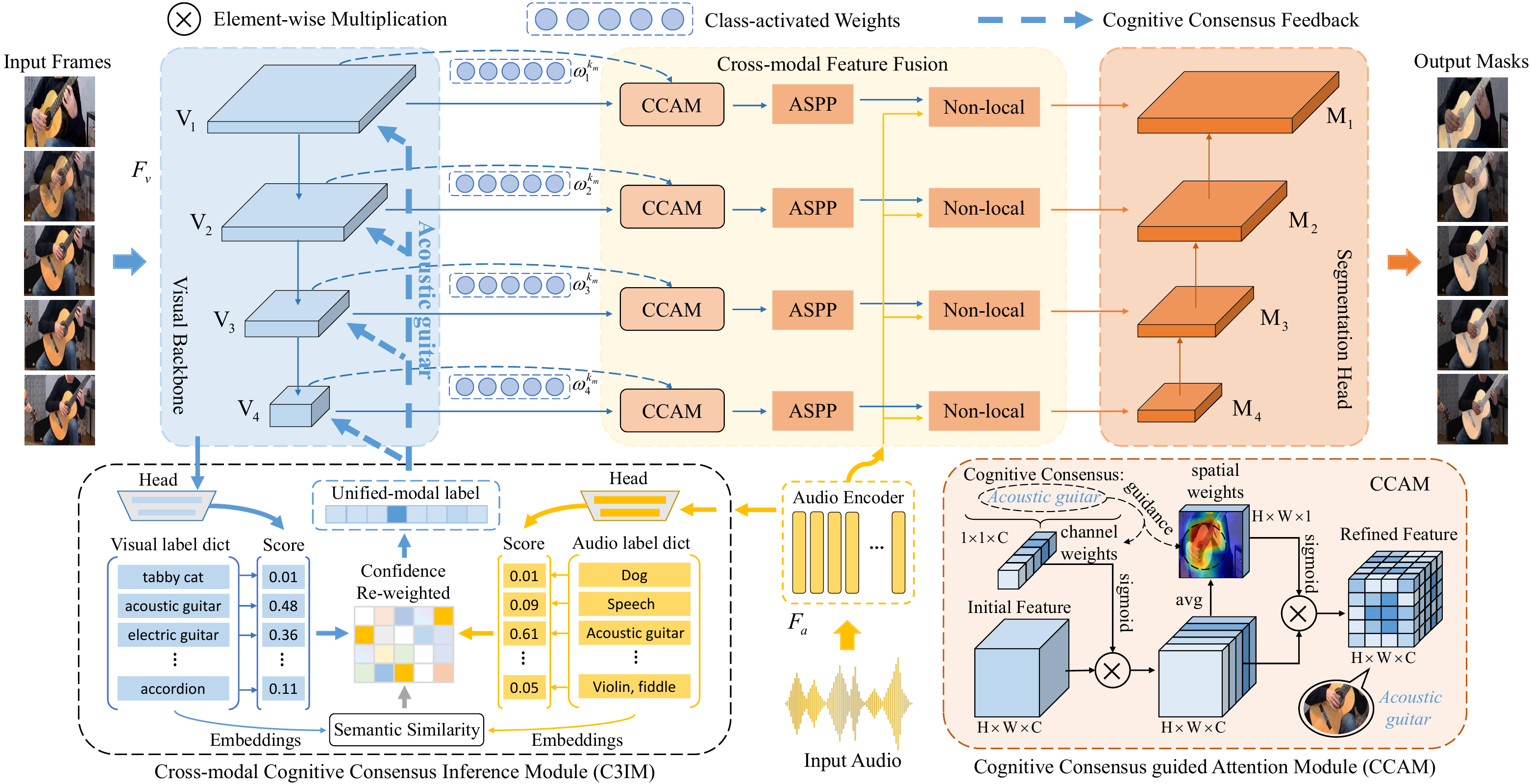}
\caption{The overview of C3N. Firstly, the audio clip $A$ and visual frames $V=\{{{I}_{t}}\}_{t=1}^{T}$ are converted into the audio feature ${{F}_{a}}$ and visual features ${{F}_{v}}=\{{{V}_{i}}\}_{i=1}^{4}$. Then, we utilize audio and visual classification confidence and the similarities between label embeddings to construct the confidence re-weighted matrix. Next, we get the unified-modal label and feed it into hierarchical layers of visual backbone to obtain class-activated weights, which guide highlighting the local feature elements through the Cognitive Consensus guided Attention Module (CCAM). Finally, a cross-modal feature fusion module composed of Atrous Spatial Pyramid Pooling (ASPP) modules and cross-modal Non-local blocks, and a segmentation head are adopted for the prediction.}
\label{fig:2}
\end{figure*}

\subsection{Label Embedding}
Word embedding means mapping distinct words into representational vectors to allow the neural networks to learn the contents and semantic relationship of words, which can be divided into static methods such as word2vec \cite{mikolov2013efficient}, Glove \cite{pennington2014glove}, and contextural-based methods such as ELMo \cite{peters2018deep}, GPT \cite{radford2018improving} and BERT \cite{devlin2018bert}. These powerful methods help researchers incorporate text semantics to facilitate the reasoning process.

In many cases, researchers convert label texts into embedding vectors and combine them with extracted features to enrich the framework's semantic-level knowledge \cite{yogatama2015embedding,zhang2017multi,wang2018joint,pappas2019gile,du2019explicit,xiong2021fusing,chen2020disentangling}. Zhang \textit{et al.} \cite{zhang2017multi} point out the importance of label information and design multiple label embedding-based models. LEAM \cite{wang2018joint} and EXAM \cite{du2019explicit} introduce interaction mechanisms to extract input-label relevance clues. GILE \cite{pappas2019gile} proposes a generalized input-label strategy to strengthen the model's performance on unseen classes. HARNN \cite{huang2019hierarchical} incorporates recurrent layers with attention mechanism and models multi-layers dependencies. Xiong \textit{et al.} \cite{xiong2021fusing}, Cai \textit{et al.} \cite{cai2020hybrid}, and Wang \textit{et al.} \cite{wang2022idea} introduce label embedding into BERT \cite{devlin2018bert} or its variants to further enhance the semantic-level sensibility of the model. LTTA-LE \cite{chen2022long} is a truncation-based method that leverages the label embedding to filter the redundant information of the long text and reduce the text length. In addition, the effectiveness of label embedding is also demonstrated in visual tasks \cite{mensink2012metric,akata2015label} and zero-shot tasks \cite{yogatama2015embedding}.

Label embedding is currently adopted mainly on the above single-modal tasks such as image classification and text classification, while exploration in global semantic-level audio-visual alignment based on label embedding is still limited.

\section{Proposed Method}
The architecture of our Cross-modal Cognitive Consensus guided Network (C3N) is illustrated in Fig. \ref{fig:2}. The network takes an audio $A$ and a video $V$ as inputs. As described in the pioneering work \cite{zhou2022audio}, the audio clips and video frames have been synchronized based on 1-second timestamps by dividing the audio into 1-second clips $A=\{{{A}_{t}}\}_{t=1}^{T}\in {{\mathbb{R}}^{T\times D}}$ and re-sampling the frames $V=\{{{I}_{\text{t}}}\}_{t=1}^{T}\in {{\mathbb{R}}^{T\times H\times W\times 3}}$ from the video with 1-second intervals, where $D$ is the number of audio signal points, $T$ is the length of the audio/video sequence and $H$ and $W$ are height and width of frames. Then, we use audio/visual encoders to extract their features. The audio features are denoted as ${{F}_{a}}=\{F_{a}^{(t)}\}_{t=1}^{T}\in {{\mathbb{R}}^{T\times d}}$, where $d$ is the feature dimension. And the visual features are ${{F}_{v}}=\{F_{v}^{(t)}\}_{t=1}^{T}$, $F_{v}^{(t)}$ contains hierarchical features $F_{v}^{(t)}=\{V_{i}^{(t)}\}_{i=1}^{4}$, where $V_{i}^{(t)}\in {{\mathbb{R}}^{{{C}_{{{v}_{i}}}}\times {{H}_{i}}\times {{W}_{i}}}}$. Then, a Cross-modal Cognitive Consensus Inference Module (C3IM) uses pre-trained heads to compute the audio/visual classification confidence and calculate the similarities between audio and visual labels for extracting the semantic-level unified-modal label. Next, the Cognitive Consensus guided Attention Module (CCAM) highlights the local feature elements of the sounding object with the guidance of the inferred unified-modal label. Finally, we feed the refined visual features into the cross-modal feature fusion module and make predictions via the segmentation head.

\subsection{Cross-modal Cognitive Consensus Inference Module}
We aim to explicitly extract global semantic-level alignment information to improve the performance of the AVS framework. Therefore, we propose a novel Cross-modal Cognitive Consensus Inference Module (C3IM) to exploit semantic-level audio/visual classification confidence and labels to infer the audio-visual cognitive consensus. The specific audio-visual semantic alignment mechanism is that the C3IM first conducts modality-specific multi-label classification to extract potential audio and visual elements with their respective classification confidence scores from the input audio and video clips. Second, the audio and visual labels are projected into a unified embedding space to calculate the modality-agnostic label similarities, which facilitates cross-modal semantic-level alignment and overcoming the appearance diversity of objects with identical semantics in complex scenarios. Finally, the C3IM integrates the classification confidence scores and label similarities to obtain the confidence re-weighted matrix, which is used to extract the audio-visual label pair of the sounding object with high semantic consistency and confidence. Since the audio and visual elements are projected into the unified semantic embedding space, the cross-modal alignment of C3IM is not affected by the number of elements.

\begin{figure}[!t]
\centering
\includegraphics[width=1.0\linewidth]{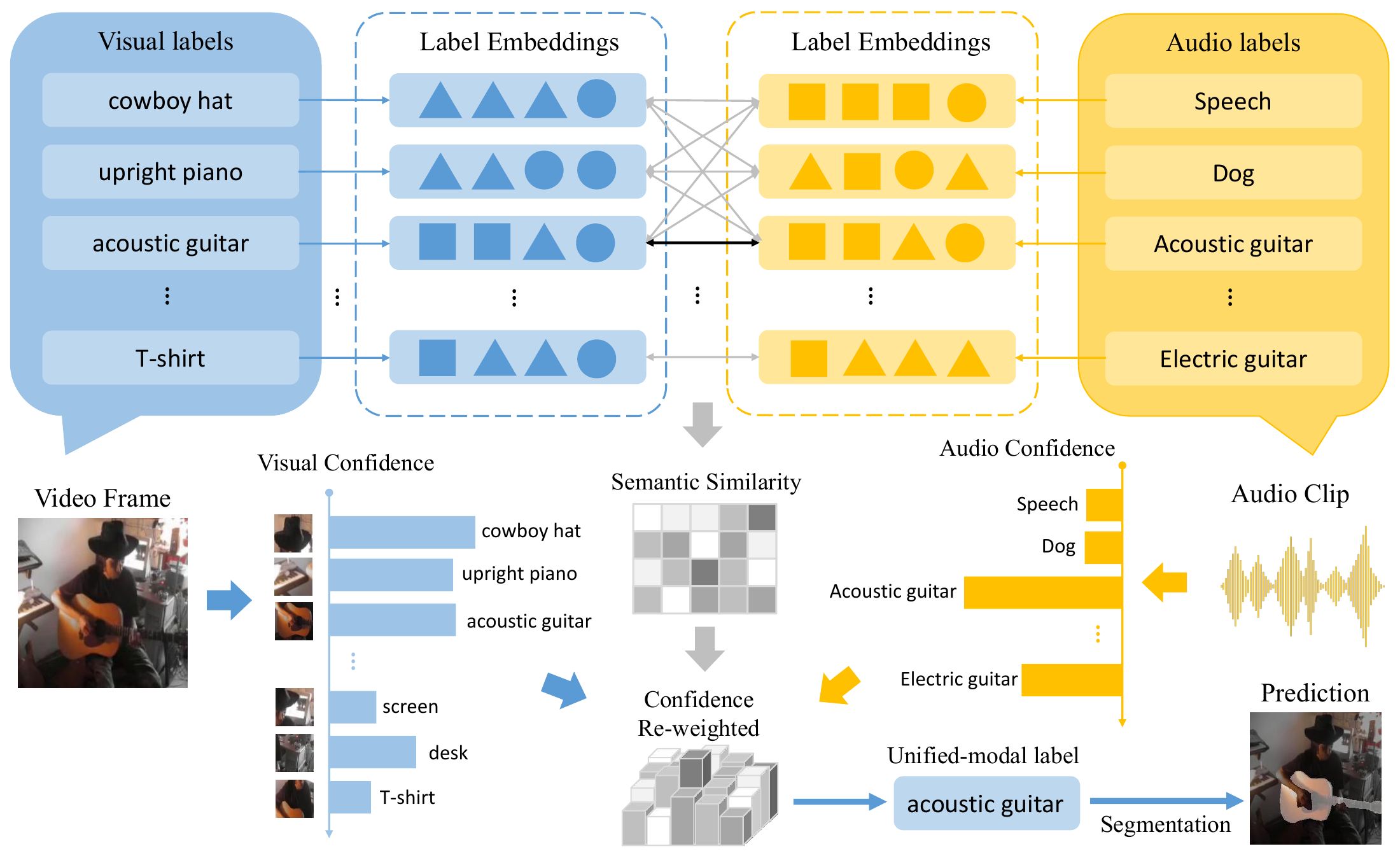}
\caption{Schematic of the C3IM. We first utilize pre-trained heads to obtain the audio/visual classification confidence ${{C}^{A}}$, ${{C}^{V}}$ independently. Then, we calculate similarities between modality-specific labels to construct the semantic similarity matrix ${{M}_{sim}}$. Finally, multiplying ${{C}^{A}}$, ${{C}^{V}}$, and ${{M}_{sim}}$ to obtain the confidence re-weighted matrix ${{M}_{cof}}$ and infer the cognitive consensus-based unified-modal label.}
\label{fig:3}
\vspace{-1mm}
\end{figure}

The visual backbone is initialized with weights pre-trained on ImageNet \cite{russakovsky2015imagenet}, a large-scale image classification dataset totaling ${N}_{v}=1000$ classes. And the audio encoder is pre-trained on AudioSet \cite{gemmeke2017audio} with ${N}_{a}=527$ audio classes. Such a diversity of classes applies to the semantic description of audio/visual objects in most scenarios. In practice, we freeze the parameters of the audio and visual classification heads. Firstly, the audio features can be represented as ${{F}_{a}}=\{F_{a}^{(1)},\cdots ,F_{a}^{(T)}\}$, where the superscripts denote timestamps. For the visual feature $F_{v}^{(t)}=\{V_{i}^{(t)}\}_{i=1}^{4}$, where $V_{i}^{(t)}$ denotes the stage $i$ visual feature of the $t$-$th$ frame. As shown in Fig. \ref{fig:3}, for every-second audio feature $F_{a}^{(t)}\in {{\mathbb{R}}^{d}}$ and the highest level visual feature $V_{4}^{(t)}\in {{\mathbb{R}}^{{{C}_{{{v}_{4}}}}\times {{H}_{4}}\times {{W}_{4}}}}$, we compute the corresponding classification confidence scores:
\begin{equation}
{{C}^{A}}=\text{Softmax}(W_{c}^{A}{F_{a}^{(t)}})
\end{equation}
\begin{equation}
{{C}^{V}}=\text{Softmax}(W_{c}^{V}\text{Avgpool}({V_{4}^{(t)}}))
\end{equation}
where $W_{c}^{A}$ and $W_{c}^{V}$ are pre-trained weights of the audio and visual classification heads, respectively. ${{C}^{A}}=\{c_{j}^{A}\}_{j=1}^{{N}_{a}}\in {{\mathbb{R}}^{{N}_{a}}}$ and ${{C}^{V}}=\{c_{k}^{V}\}_{k=1}^{{N}_{v}}\in {{\mathbb{R}}^{{N}_{v}}}$ denote classification confidence scores, where $c_{j}^{A}$ and $c_{k}^{V}$ are confidence of the $j\text{-}th$ audio class and $k\text{-}th$ visual class.

Then, to bridge the semantic gap between distinct modalities, we measure the similarities of audio and visual labels and construct a semantic similarity matrix. The AudioSet and ImageNet labels are denoted as ${{L}^{A}}=\{l_{j}^{A}\}_{j=1}^{{N}_{a}}$ and ${{L}^{V}}=\{l_{k}^{V}\}_{k=1}^{{N}_{v}}$ respectively. We convert the audio and visual labels into a unified label embedding space through SpaCy, an advanced library for natural language processing to calculate similarities between label words or phrases, as follows:
\begin{equation}
{{E}^{A}}=\mathcal{E}({{L}^{A}})
\end{equation}
\begin{equation}
{{E}^{V}}=\mathcal{E}({{L}^{V}})
\end{equation}
where $\mathcal{E}(\cdot)$ denotes the embedding layer. The audio and visual label embeddings are represented as ${{E}^{A}}=\{e_{j}^{A}\}_{j=1}^{{N}_{a}}\in {{\mathbb{R}}^{{N}_{a}\times {d}'}}$ and ${{E}^{V}}=\{e_{k}^{V}\}_{k=1}^{{N}_{v}}\in {{\mathbb{R}}^{{N}_{v}\times {d}'}}$ respectively, where the embedding dimension ${d}'$ is 300. The similarity between the audio and visual labels is defined by the following equation:
\begin{equation}
{{m}_{j,k}^{sim}}=\frac{e_{j}^{A}\cdot {{(e_{k}^{V})}^{T}}}{||e_{j}^{A}|{{|}_{F}}||e_{k}^{V}|{{|}_{F}}}
\end{equation}
where $||\cdot |{{|}_{F}}$ denotes the Frobenius norm, ${m}_{j,k}^{sim}$ is an element of the semantic similarity matrix ${{M}_{sim}}\in {{\mathbb{R}}^{{N}_{a}\times {N}_{v}}}$, and $j$, $k$ denote the row and column indexes.

Finally, based on the computed audio and visual classification confidence scores ${{C}^{A}}=\{c_{j}^{A}\}_{j=1}^{{N}_{a}}$, ${{C}^{V}}=\{c_{k}^{V}\}_{k=1}^{{N}_{v}}$ and the similarity matrix ${{M}_{sim}}$, a confidence re-weighted matrix ${{M}_{cof}}\in {{\mathbb{R}}^{{N}_{a}\times {N}_{v}}}$ can be calculated by the following equation:
\begin{equation}
{{M}_{cof}}(j,k)={{(c_{j}^{A})}^{\alpha }}\cdot {{M}_{sim}}(j,k)\cdot {{(c_{k}^{V})}^{\beta }}
\label{eq:(4)}
\end{equation}
where $\alpha $ and $\beta $ are balance coefficients, which are set to 0.1. Values in the re-weighted matrix ${{M}_{cof}}$ can also be considered as cross-modal cognitive consensus degrees. With the inferred cognitive consensus, semantically relevant audio-visual objects can be identified to guide the following segmentation. For ambiguous or noisy audio-visual environments, on the one hand, the unified-modal labels are inferred from the integration and complementarity of the audio and visual modality information, which improves the model's robustness when a single modality is ambiguous or noisy. On the other hand, previous works \cite{ulyanov2018deep,cheng2019bayesian} have proven that the structures of deep neural networks are able to capture the input data's statistical prior information and prevent overfitting to the noise.

\subsection{Cognitive Consensus guided Attention Module}
For ease of representation, the time superscript $(t)$ is omitted in this and subsequent sections. After the cross-modal cognitive consensus inference, we feed the semantically consistent unified-modal label to the hierarchical layers of the visual backbone. In addition, we propose a parameter-free Cognitive Consensus guided Attention Module (CCAM), which injects class-activated weights into visual features to highlight the local feature elements corresponding to the sounding object.

We search for the maximum value $m_{jm,km}^{cog}$ of ${{M}_{cof}}$ and obtain its row and column indexes ${{j}_{m}}$ and ${{k}_{m}}$:
\begin{equation}
m_{jm,km}^{cog}=\max\limits_{\substack{1\le j\le {N}_{a} \\ 1\le k\le {N}_{v}}}{{M}_{cog}}(j,k)
\end{equation}
Specifically, ${{j}_{m}}$ and ${{k}_{m}}$ refer to the index of the audio and visual classes with the highest score. Inspired by the ideology of Grad-CAM \cite{selvaraju2017grad,chattopadhay2018grad}, we calculate the loss for the ${{k}_{m}}\text{-}th$ visual class and perform the backpropagation to compute the corresponding gradients. Different from Grad-CAM, which computes class-activated heat maps, we aim to obtain the weights corresponding to the unified-modal label with the semantic-level cognitive consensus as follows:
\begin{equation}
\omega _{z}^{{{k}_{m}}}=\frac{1}{H\times W}\sum\limits_{x}{\sum\limits_{y}{\frac{\partial {{y}^{{{k}_{m}}}}}{\partial A_{xy}^{z}}}}
\end{equation}
where ${{y}^{{{k}_{m}}}}$ denotes classification logit of the ${{k}_{m}}\text{-}th$ visual class, $A$ denotes feature map activations, $z$ denotes the index of the channel, $x,y$ denotes the spatial coordinates of the feature map. In practice, we take the last layer of each stage as the activation layer and the class-activated weights of stage $i$ are denoted as $\omega _{i}^{{{k}_{m}}}=\{\omega _{i,z}^{{{k}_{m}}}\}_{z=1}^{{{C}_{{{v}_{i}}}}}$, and $\omega _{i}^{{{k}_{m}}}\in {{\mathbb{R}}^{{{C}_{{{v}_{i}}}}}}$. 

To integrate the class-activated weights $\omega _{i}^{{{k}_{m}}}$ with the initial visual features ${{V}_{i}}$ for highlighting the local feature elements corresponding to the cognitive consensus, we propose a parameter-free CCAM, whose structure is shown in the bottom right of Fig \ref{fig:2}. Without any learnable parameters, CCAM performs channel-spatial attention to the initial visual features and outputs refined features $V_{i}^{r}$ with the guidance of inferred global semantic-level unified-modal label, as follows:
\begin{equation}
\omega _{i}^{cha}=\sigma (\omega _{i}^{{{k}_{m}}})
\end{equation}
\begin{equation}
V_{i}^{c}={{V}_{i}}\otimes \omega _{i}^{cha}
\end{equation}
\begin{equation}
\omega _{i}^{spa}=\sigma (\text{cAvg}(V_{i}^{c}))
\end{equation}
\begin{equation}
V_{i}^{r}=V_{i}^{c}\otimes \omega _{i}^{spa}
\end{equation}
where $\sigma (\cdot )$ denotes the sigmoid function, $\otimes $ denotes element-wise multiplication with broadcast mechanism, and $\text{cAvg}$ means performing the average operation on the feature along the channel. By means of the channel-spatial attention mechanism, CCAM incorporates semantic-level cognitive consensus weights with feature-level visual representations to guide the model focus on the semantically consistent visual object.

\begin{figure}[t]
\centering
\includegraphics[width=1.0\linewidth]{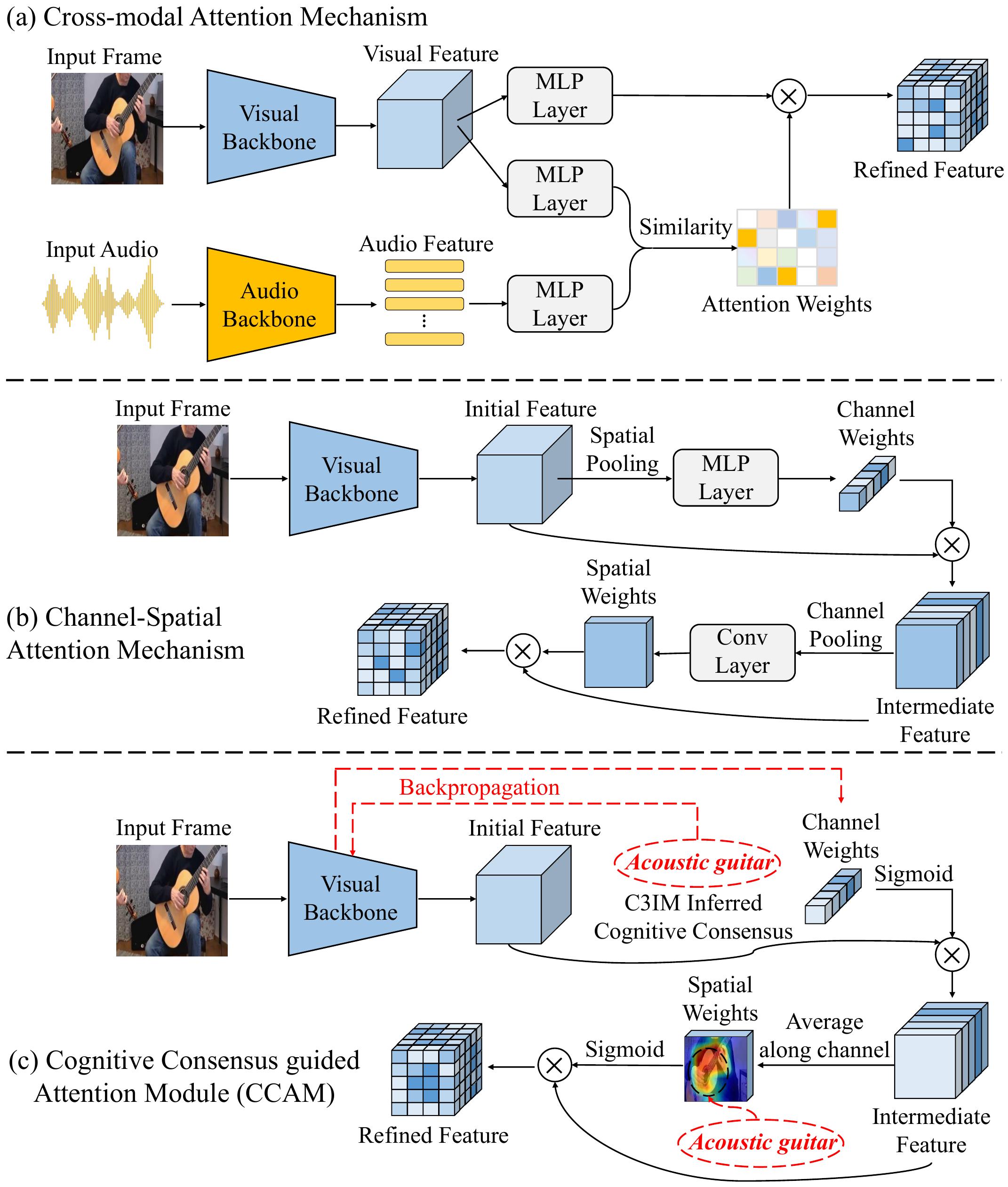}
\caption{The implementation of the commonly used cross-modal attention mechanism (shown in panel (a)), channel-spatial attention mechanism (shown in panel (b)), and our Cognitive Consensus guided Attention Module (CCAM) (shown in panel (c)).}
\label{fig:8}
\end{figure}

As shown in Fig. \ref{fig:8}, although CCAM adopts the well-explored channel-spatial attention mechanism operation similar to the panel (b) of Fig. \ref{fig:8}, CCAM utilizes the cognitive consensus inferred by C3IM with explicit semantic information and performs backpropagation to obtain the class-activated weights $\omega _{i}^{{{k}_{m}}}$, and they are then injected into the visual features to highlight the audio-visual semantic consistent object in a channel-spatial attention manner as in the panel (c) of Fig. \ref{fig:8}. Unlike the forward propagate \textit{feature-level} similarity-based cross-modal attention modules \cite{zhou2022audio,hao2023improving} as shown in the panel (a) of Fig. \ref{fig:8}, our CCAM relies on the attention weights with audio-visual \textit{semantic-level} alignment information obtained by backpropagation operation.

\subsection{Cross-modal Feature Fusion}
For cognitive consensus refined visual features $V_{i}^{r}\in {{\mathbb{R}}^{T\times {{C}_{{{v}_{i}}}}\times {{H}_{i}}\times {{W}_{i}}}}$, Atrous Spatial Pyramid Pooling (ASPP) \cite{chen2017deeplab} modules are used to help the perception at multiple receptive fields. We denote the ASPP-processed features as $V_{i}^{a}\in {{\mathbb{R}}^{T\times C\times {{H}_{i}}\times {{W}_{i}}}}$.

After getting the visual feature $V_{i}^{a}$, feed it with the audio feature ${{F}_{a}}$ into a cross-modal Non-local block \cite{wang2018non,zhou2022audio}, which utilizes 3-D convolutional layers to model the feature-level temporal-spatial dependencies and captures the inter-modal correlations. The cross-modal Non-local block-based feature fusion performs dense feature-level interaction to calibrate the visual features via the discriminative audio features, which enforces highlighting the visual context of sounding objects and ensures accurate segmentation. We first map and repeat ${{F}_{a}}$ into ${{\hat{F}}_{a}}\in {{\mathbb{R}}^{T\times C\times {{H}_{i}}\times {{W}_{i}}}}$. Then, the cross-modal feature-level interaction process at stage $i$ can be formulated as follows:
\begin{equation}
\Phi =\frac{{{\theta }_{1}}\left( V_{i}^{a} \right)\cdot {{\theta }_{2}}{{({{{\hat{F}}}_{a}})}^{T}}}{N}
\end{equation}
\begin{equation}
{{M}_{i}}=V_{i}^{a}+{{\theta }_{4}}(\Phi \cdot {{\theta }_{3}}(V_{i}^{a}))
\end{equation}
where ${{\theta }_{1}}$, ${{\theta }_{2}}$, ${{\theta }_{3}}$ and ${{\theta }_{4}}$ denote $1\times 1\times 1$ 3-D convolutional layers. $N$ is the number of pixels of the feature map, which is treated as a normalization constant. $\Phi $ is the cross-modal attention matrix, which measures the pixel-wise correlations between audio and visual features. ${{M}_{i}}$ denotes the multi-modal feature of the $i\text{-}th$ stage.

\subsection{Segmentation Head}
We combine the hierarchical multi-modal features $\{{{M}_{i}}\}_{i=1}^{4}$ in a simple top-down manner by progressively upsampling the higher-level feature and integrating it with the lower-level feature rich in detail information. This process can be formulated as follow:
\begin{equation}
\left\{ \begin{array}{*{35}{cr}}
   {{Y}_{i}}=\text{Upsample}(\text{Conv}({{Y}_{i+1}}\text{+Conv}({{M}_{i}})))&\text{        1}\le i < \text{4}  \\
   {{Y}_{i}}=\text{Upsample}(\text{Conv}({{M}_{i}})) \qquad  & i=4  \\
\end{array} \right.
\end{equation}
Then the final prediction masks $\hat{Y}=\{{{\hat{Y}}_{t}}\}_{t=1}^{T}\in {{\mathbb{R}}^{T\times H\times W}}$ can be obtained by feeding ${{Y}_{1}}$ into a fully convolutional head.

To optimize the parameters of the proposed model. We adopt the BCE loss and Dice loss as the loss function:
\begin{equation}
{{L}_{seg}}=BCELoss(\hat{Y},Y)+DiceLoss(\hat{Y},Y)
\end{equation}
where $Y\in {{\mathbb{R}}^{T\times H\times W}}$ denotes the ground truth masks.

\section{Experiments}
\subsection{Experimental Settings}
\subsubsection{Dataset}
The proposed method is evaluated on the mainstream AVSbench \cite{zhou2022audio} dataset. For the construction of the AVSBench dataset, Zhou \textit{et al.} \cite{zhou2022audio} first collect the videos from YouTube using the technique in VGGSound \cite{chen2020vggsound} to guarantee the audio-visual intended semantics. Then, depending on the number of sounding objects in the video, the AVSBench dataset is divided into a Single-Source subset and a Multi-Source subset, which are further split into train/val/test sets. Finally, the video frames are annotated with the binary pixel-level masks for the sounding object with 1-second intervals. In practice, each video of AVSBench is cropped to 5 seconds, and video frames at the end of each second are extracted. The Single-Source subset contains 4,392 videos from 23 categories. Note that only the first sampled frame is annotated for videos in the train split, while all sampled frames of the val/test split are annotated, totaling 10,852 annotations. The Multi-Sources subset contains 424 videos and 2,120 annotated pixel-level segmentation masks. All sampled frames of the train/val/test split have annotation since the sounding objects may change over time. As with the pioneering work \cite{zhou2022audio}, the following experiments are under the semi-supervised Single Sound Source Segmentation (S4) and the fully supervised Multiple Sound Source Segmentation (MS3) settings.

\subsubsection{Implementation Details}
We implement the proposed method using PyTorch \cite{paszke2019pytorch}. We use Swin Transformer (Swin) \cite{liu2021swin} and Pyramid Vision Transformer v2 (PVTv2) \cite{wang2022pvt} as the visual backbones. The channel sizes of the four stages are ${{C}_{{{v}_{1}}:{{v}_{4}}}}=\{128,256,512,1024\}$ for Swin and ${{C}_{{{v}_{1}}:{{v}_{4}}}}=\{64,128,320,512\}$ for PVTv2. The visual backbones, including the corresponding classification heads, are pre-trained on ImageNet-1K \cite{russakovsky2015imagenet} dataset and the weights of the classification heads are frozen. For audio encoders, the widely adopted VGGish \cite{hershey2017cnn}, PANNs \cite{kong2020panns}, and BEATs \cite{chen2022beats} pre-trained on AudioSet \cite{gemmeke2017audio} are used to extract audio features. Since the VGGish does not have a classification head, we trained an additional head for classification using audio features from AudioSet. All of the video frames are resized to $224\times 224$, and the audio clips are clipped to 1-second splits. Due to the limitation of the data scale of AVSBench, we perform the ColorJitter and RandomHorizonFlip data augmentation strategy. The channel size of $C$ is set to 256. In all experiments, the models are optimized by Adam \cite{kingma2014adam} and the initial learning rate is 0.0001. The batch size is set to 8 and the number of training epochs is 20 for the S4 setting. For the MS3 setting, the batchsize is 4 and the number of training epochs is 60.

\subsubsection{Evaluate Metrics}
We adopt the mean intersection-over-union (mIoU) and F-score as evaluation metrics to measure the performance of the proposed method quantitatively. The mIoU is calculated by dividing the intersection area by the union area of the predictions and ground truths and taking the average value of the whole dataset and F-score \footnote{${F}=\frac{(1+{{\beta }^{2}})\times precision\times recall}{{{\beta }^{2}}\times precision+recall}$, ${\beta }^{2}$ is set to 0.3, which remains the same as in \cite{zhou2022audio}.} considers both precision and recall.

\subsection{Comparison with State-of-the-Art Methods}
\begin{table}[t]
\centering
\caption{Quantitative results under the S4 and MS3 setting of the mIoU and F-score metrics (TPAVI-Re and ECMVAE-Re means the results of TPAVI \cite{zhou2022audio} and ECMVAE \cite{mao2023multimodal} that we re-implemented).}
\label{tab:1}
{
\begin{tabular}{c|l|c|c|c|c|c|c}
\toprule
\multicolumn{2}{l|}{\multirow{2}{*}{Method}} &

\multicolumn{2}{l|}{\multirow{2}{*}{Backbone}} &

\multicolumn{2}{c|}{mIoU} & \multicolumn{2}{c}{F-score}
 \\ \cline{5-8} 
\multicolumn{2}{c|}{}                         &
\multicolumn{2}{c|}{}                         &
\multicolumn{1}{c|}{S4} & \multicolumn{1}{c|}{MS3} & \multicolumn{1}{c|}{S4} & \multicolumn{1}{c}{MS3}     \\ \hline \hline
\multicolumn{2}{l|}{LVS\cite{chen2021localizing}} &  \multicolumn{2}{l|}{ResNet-18 $\times$ 2} & 37.94  & 29.45  & 0.510  & 0.330  \\ 
\multicolumn{2}{l|}{MSSL\cite{qian2020multiple}} &  \multicolumn{2}{l|}{ResNet-18+CRNN} & 44.89  & 26.13  & 0.663  & 0.363  \\ 
\multicolumn{2}{l|}{3DC\cite{mahadevan2020making}} &  \multicolumn{2}{l|}{ResNet-152 (3D)} & 57.10  & 36.92  & 0.759  & 0.503  \\ 
\multicolumn{2}{l|}{SST\cite{duke2021sstvos}} &  \multicolumn{2}{l|}{ResNet-101} & 66.29  & 42.57  & 0.801  & 0.572  \\ 
\multicolumn{2}{l|}{iGAN\cite{mao2021transformer}} &  \multicolumn{2}{l|}{Swin} & 61.59  & 42.89  & 0.778  & 0.544  \\ 
\multicolumn{2}{l|}{LGVT\cite{zhang2021learning}} &  \multicolumn{2}{l|}{Swin} & 74.94  & 40.71  & 0.873  & 0.593  \\ 
\hline \hline 
\multicolumn{2}{l|}{TPAVI\cite{zhou2022audio}} &  \multicolumn{2}{l|}{ResNet-50+VGGish} & 72.79  & 47.88  & 0.848  & 0.578  \\
\multicolumn{2}{l|}{TPAVI\cite{zhou2022audio}} &  \multicolumn{2}{l|}{PVTv2+VGGish} & 78.74  & 54.00  & 0.879  & 0.645  \\ 
\multicolumn{2}{l|}{TPAVI-Re} &  \multicolumn{2}{l|}{ResNet-50+VGGish} & 72.69  & 46.56  & 0.832  & 0.563  \\
\multicolumn{2}{l|}{TPAVI-Re} &  \multicolumn{2}{l|}{ResNet-50+PANNs} & 73.06  & 48.36 &  0.834 & 0.588  \\
\multicolumn{2}{l|}{TPAVI-Re} &  \multicolumn{2}{l|}{PVTv2+VGGish} & 78.78  & 54.00  & 0.878  & 0.651  \\ 
\multicolumn{2}{l|}{TPAVI-Re} &  \multicolumn{2}{l|}{PVTv2+PANNs} &  78.98  & 55.13 & 0.879 &  0.663 \\ 
\multicolumn{2}{l|}{AVSC\cite{liu2023audiovisual}} &  \multicolumn{2}{l|}{PVTv2+BEATs} & 81.29  & 59.50 & 0.886  &  0.657\\
\multicolumn{2}{l|}{LDM\cite{mao2023contrastive}} &  \multicolumn{2}{l|}{PVTv2+VGGish} & 81.38 & 58.18 & 0.902  &  0.709 \\
\multicolumn{2}{l|}{BG\cite{hao2023improving}} &  \multicolumn{2}{l|}{PVTv2+VGGish} & 81.71 & 55.10 & 0.904  &  0.668 \\
\multicolumn{2}{l|}{ECMVAE\cite{mao2023multimodal}} &  \multicolumn{2}{l|}{PVTv2+VGGish} & 81.74 & 57.84 &  0.901 &  0.708 \\
\multicolumn{2}{l|}{ECMVAE-Re} &  \multicolumn{2}{l|}{PVTv2+VGGish} & 81.64 & 57.44 & 0.899 & 0.690  \\
\multicolumn{2}{l|}{ECMVAE-Re} &  \multicolumn{2}{l|}{PVTv2+PANNs} & 81.93 & 58.04 & 0.900  &  0.708 \\
\multicolumn{2}{l|}{BAVS\cite{liu2023bavs}} &  \multicolumn{2}{l|}{PVTv2+BEATs} & 82.68 & 59.63 &  0.898 &  0.659 \\
\hline \hline 
\multicolumn{2}{l|}{C3N (Ours)} &  \multicolumn{2}{l|}{Swin+VGGish} & 81.63 & 56.38  & 0.897 & 0.663  \\
\multicolumn{2}{l|}{C3N (Ours)} &  \multicolumn{2}{l|}{Swin+PANNs} & 81.87  & 57.09  & 0.898  & 0.671  \\
\multicolumn{2}{l|}{C3N (Ours)} &  \multicolumn{2}{l|}{Swin+BEATs} & 81.95  & 58.14 & 0.900 &  0.683 \\
\multicolumn{2}{l|}{C3N (Ours)} &  \multicolumn{2}{l|}{PVTv2+VGGish} & 82.67  & 60.11  & 0.904  & 0.697 \\
\multicolumn{2}{l|}{C3N (Ours)} &  \multicolumn{2}{l|}{PVTv2+PANNs} &  82.94 &  61.67  & 0.906  &  0.713 \\
\multicolumn{2}{l|}{C3N (Ours)} &  \multicolumn{2}{l|}{PVTv2+BEATs} & \textbf{83.11} & \textbf{61.72} & \textbf{0.908} & \textbf{0.722}  \\
\bottomrule
\end{tabular}
}
\end{table}

We use the mIoU and F-score metrics to quantitively compare our method with state-of-the-art methods \cite{chen2021localizing,qian2020multiple,mahadevan2020making,duke2021sstvos,mao2021transformer,zhang2021learning,zhou2022audio,liu2023audiovisual,mao2023contrastive,hao2023improving,mao2023multimodal,liu2023bavs} under the S4 and MS3 settings. We show the performance of SSL/VOS/Salient Object Detection (SOD) methods on AVSBench reported in \cite{zhou2022audio} in the first panel of Table \ref{tab:1}. In addition, in the second panel, we also show the performance of many brand-new AVS methods for comprehensive comparison, and we re-implement the official source code of TPAVI \cite{zhou2022audio} and ECMVAE \cite{mao2023multimodal}. In addition, we also re-implement them with PANNs \cite{kong2020panns} as the audio backbone for a fair assessment. Finally, in the third panel, we show the performance of our C3N under six different backbone combinations.

Table \ref{tab:1} shows the results of the test set. The proposed C3N with PVTv2 \cite{wang2022pvt} as the visual backbone and VGGish \cite{hershey2017cnn} or PANNs \cite{kong2020panns}  as the audio encoder achieves remarkable performance and outperforms other AVS methods. For the simple S4 setup, compared with BAVS \cite{liu2023bavs} with PVTv2 as the visual backbone and strong BEATs \cite{chen2022beats} as the audio backbone, our C3N (PVTv2+PANNs) gains higher mIoU and F-score and C3N (PVTv2+VGGish) achieves comparable performance. For the difficult MS3 setting with multiple sounding objects, thanks to the guidance of semantic-level cognitive consensus, C3N (PVTv2+VGGish) and C3N (PVTv2+PANNs) achieve higher performances than BAVS by 0.48$\%$, 2.04$\%$ in mIoU and 0.038, 0.054 in F-score, respectively. 

We also adopt Swin Transformer \cite{liu2021swin} to evaluate the general effectiveness of our cognitive consensus-based method. Our C3N (Swin+VGGish) achieves 81.63$\%$ mIoU and 0.897 F-score under the S4 setting and 56.38$\%$ mIoU and 0.663 F-score under the MS3 setting. While the results of C3N (Swin+PANNs) are higher. Despite the PVTv2 backbone is stronger than the Swin backbone, the performance of C3N with Swin as the backbone is comparable with some PVTv2-based AVS methods and far higher than the classical TPAVI (ResNet version) under the S4 and MS3 settings. In addition, we also adopt BEATs \cite{chen2022beats} as the audio backbone, which leads to mIoU and F-score performance improvement compared with C3N with the VGGish or PANNs as the audio backbone under the S4 and MS3 settings. Specifically, our C3N (Swin+BEATs) outperforms C3N (Swin+PANNs) by 1.05$\%$ in mIoU and 0.012 in F-score under the MS3 setting and achieves slightly higher mIoU and F-score under the S4 setting. C3N (PVTv2+BEATs) achieves higher performance than C3N (PVTv2+PANNs) by 0.05$\%$ in mIoU and 0.009 in F-score under the MS3 setting, and 0.17$\%$ in mIoU and 0.002 in F-score under the S4 setting.

\subsection{Ablation Study}
\subsubsection{Analysis of key components}
\begin{table}[t]
\centering
\caption{Ablation results on the test set of AVSBench under the S4 and MS3 settings. The results under all four backbone settings are presented. AF denotes audio features, CC-V denotes the cognitive consensus relies only on the visual modality, and CC denotes the proposed audio-visual cognitive consensus.}
\label{tab:2}
\setlength{\tabcolsep}{2.0mm}
{
\begin{tabular}{c|c|c|c|c|c|c}
\toprule
\multicolumn{7}{c}{Swin+VGGish}\\  \hline \hline
     \multicolumn{1}{c}{\multirow{2}{*}{AF}}     & 
     \multicolumn{1}{c}{\multirow{2}{*}{CC-V}}   &
     \multicolumn{1}{c|}{\multirow{2}{*}{CC}}     &
     \multicolumn{2}{c|}{mIoU}                     & \multicolumn{2}{c}{F-score} \\ \cline{4-7}
  \multicolumn{1}{l}{}  & \multicolumn{1}{l}{} & \multicolumn{1}{l|}{} & \multicolumn{1}{c|}{S4}              & \multicolumn{1}{c|}{MS3} & \multicolumn{1}{c|}{S4}              & \multicolumn{1}{c}{MS3}\\ \hline
\multicolumn{1}{l}{}  &\multicolumn{1}{l}{}  & \multicolumn{1}{l|}{}  & 79.07 &  48.81 & 0.878  &  0.583\\
\multicolumn{1}{l}{}  &\multicolumn{1}{l}{\checkmark}  & \multicolumn{1}{l|}{}  & 80.02  &  49.38  & 0.882  & 0.589 \\
\multicolumn{1}{l}{}  &\multicolumn{1}{l}{}  &\multicolumn{1}{l|}{\checkmark}  & 80.37  &  50.32  & 0.887   & 0.609 \\
\multicolumn{1}{l}{\checkmark}  &\multicolumn{1}{l}{}  &\multicolumn{1}{l|}{}  & 80.61  &  53.05  & 0.889  &  0.625\\
\multicolumn{1}{l}{\checkmark}  &\multicolumn{1}{l}{}  & \multicolumn{1}{l|}{\checkmark}  & \textbf{81.63}  &  \textbf{56.38}  &  \textbf{0.897}  & \textbf{0.663} \\ 
\bottomrule
\toprule
\multicolumn{7}{c}{Swin+PANNs}\\  \hline \hline
     \multicolumn{1}{c}{\multirow{2}{*}{AF}}     & \multicolumn{1}{c}{\multirow{2}{*}{CC-V}}     & 
     \multicolumn{1}{c|}{\multirow{2}{*}{CC}}     &
     \multicolumn{2}{c|}{mIoU}                     & \multicolumn{2}{c}{F-score} \\ \cline{4-7}
  \multicolumn{1}{l}{}  & \multicolumn{1}{l}{} & \multicolumn{1}{l|}{} & \multicolumn{1}{c|}{S4}              & \multicolumn{1}{c|}{MS3} & \multicolumn{1}{c|}{S4}              & \multicolumn{1}{c}{MS3}\\ \hline
\multicolumn{1}{l}{} &\multicolumn{1}{l}{} &\multicolumn{1}{l|}{}  & 79.59  &  48.68  & 0.882  & 0.595 \\
\multicolumn{1}{l}{}  &\multicolumn{1}{l}{\checkmark}  & \multicolumn{1}{l|}{}  & 80.38  &  50.56  & 0.886  & 0.606 \\
\multicolumn{1}{l}{} &\multicolumn{1}{l}{} &\multicolumn{1}{l|}{\checkmark}  &  80.45  & 51.59  & 0.886   &  0.629\\
\multicolumn{1}{l}{\checkmark} &\multicolumn{1}{l}{} &\multicolumn{1}{l|}{}  & 80.69 & 54.34  & 0.889 & 0.649\\
\multicolumn{1}{l}{\checkmark} &\multicolumn{1}{l}{} &\multicolumn{1}{l|}{\checkmark}  &  \textbf{81.87} &  \textbf{57.09}  &  \textbf{0.898} & \textbf{0.671} \\ 
\bottomrule
\toprule
\multicolumn{7}{c}{PVTv2+VGGish}\\  \hline \hline
     \multicolumn{1}{c}{\multirow{2}{*}{AF}}     & 
     \multicolumn{1}{c}{\multirow{2}{*}{CC-V}}    &
     \multicolumn{1}{c|}{\multirow{2}{*}{CC}}     &
     \multicolumn{2}{c|}{mIoU}                     & \multicolumn{2}{c}{F-score} \\ \cline{4-7}
  \multicolumn{1}{l}{}  & \multicolumn{1}{l}{}  & \multicolumn{1}{l|}{} & \multicolumn{1}{c|}{S4}              & \multicolumn{1}{c|}{MS3} & \multicolumn{1}{c|}{S4}              & \multicolumn{1}{c}{MS3}\\ \hline
\multicolumn{1}{l}{} &\multicolumn{1}{l}{} &\multicolumn{1}{l|}{}  &  80.76 &  50.75  & 0.888   & 0.611 \\
\multicolumn{1}{l}{}  &\multicolumn{1}{l}{\checkmark}  & \multicolumn{1}{l|}{}  & 81.62 & 52.26   &  0.895 & 0.622 \\
\multicolumn{1}{l}{} &\multicolumn{1}{l}{} &\multicolumn{1}{l|}{\checkmark}  & 81.92  &  54.30  & 0.897  & 0.644\\
\multicolumn{1}{l}{\checkmark} &\multicolumn{1}{l}{} &\multicolumn{1}{l|}{}  &  81.70  &  56.73 & 0.896 & 0.663\\
\multicolumn{1}{l}{\checkmark} &\multicolumn{1}{l}{} &\multicolumn{1}{l|}{\checkmark}  & \textbf{82.67} &  \textbf{60.11} & \textbf{0.904}  & \textbf{0.697} \\ 
\bottomrule
\toprule
\multicolumn{7}{c}{PVTv2+PANNs}\\  \hline \hline
     \multicolumn{1}{c}{\multirow{2}{*}{AF}}     & 
     \multicolumn{1}{c}{\multirow{2}{*}{CC-V}}     &
     \multicolumn{1}{c|}{\multirow{2}{*}{CC}}     &
     \multicolumn{2}{c|}{mIoU}                     & \multicolumn{2}{c}{F-score} \\ \cline{4-7}
  \multicolumn{1}{l}{}  & \multicolumn{1}{l}{}  &\multicolumn{1}{l|}{} & \multicolumn{1}{c|}{S4}              & \multicolumn{1}{c|}{MS3} & \multicolumn{1}{c|}{S4}              & \multicolumn{1}{c}{MS3}\\ \hline
\multicolumn{1}{l}{} &\multicolumn{1}{l}{}  &\multicolumn{1}{l|}{}  &  80.73 &  52.18 &  0.890 & 0.617 \\
\multicolumn{1}{l}{}  &\multicolumn{1}{l}{\checkmark}  & \multicolumn{1}{l|}{}  & 81.40 & 53.57   & 0.894  & 0.634 \\
\multicolumn{1}{l}{} &\multicolumn{1}{l}{}  &\multicolumn{1}{l|}{\checkmark}  &  81.97 &  54.08 &  0.896  & 0.651\\
\multicolumn{1}{l}{\checkmark} &\multicolumn{1}{l}{}  &\multicolumn{1}{l|}{}  & 82.06 &   58.76 & 0.897  & 0.687 \\
\multicolumn{1}{l}{\checkmark} &\multicolumn{1}{l}{}  &\multicolumn{1}{l|}{\checkmark}  &  \textbf{82.94} &  \textbf{61.67}  & \textbf{0.906}   & \textbf{0.713} \\ 
\bottomrule

\end{tabular}
}
\end{table}

In Table \ref{tab:2}, we conduct ablation experiments under all four backbone combinations. In the experiments, we investigate the effects of audio-visual feature-level interaction, cognitive consensus relies merely on visual modality, and the full audio-visual cognitive consensus.

The first row shows the performance without audio features and cognitive consensus, i.e. with only video frames as input. Under all four backbone combinations, as in \cite{zhou2022audio}, visual-only input does not lead to a performance drop by a large margin for the S4 setting while causing a distinct performance drop under the MS3 setting. It indicates that for videos with multiple sounding sources, the introduction of the audio signal and alignment between audio-visual modalities are especially essential to the final prediction. Moreover, in the second row, we introduce the ``CC-V", which means that the labels with the highest visual semantic confidence are fed into the framework. The results show that the performance after the introduction of ``CC-V" is improved. However, relying only on semantic-level guidance from a single modality brings limited improvement.

Then we add semantic-level cognitive consensus in the third row. It means the audio-visual semantic-level cognitive consensus is introduced but does not yet include audio features. The audio-visual semantic-level cognitive consensus leads to performance gains for all four groups. In detail, under the MS3 setting, mIoU increases by 1.51$\%$, 2.91$\%$, 3.55$\%$, and 1.90$\%$ respectively, and F-score increases between 0.026 to 0.034. For the S4 setting, mIoU increases by around 1$\%$ under each backbone combination, and the F-score also increases by various points. Furthermore, compared with the results in the fourth row that add audio features, the results in the third row are competitive under the S4 setting. The above results demonstrate that the model performance can be improved through the cross-modal cognitive consensus-based method.

We incorporate audio features and conduct feature-level interaction in the fourth row. The results represent that introducing audio features to the visual input is effective. The reason is that powerful pre-trained audio encoders can capture the dense information of the audio signal. Moreover, the non-local blocks extract comprehensive cross-modal pixel-wise correlations, which is beneficial to the integration of the two modalities. Nevertheless, such cross-modal interaction is inexplicit due to the absence of semantic-level cognitive consensus, which leads to some errors during the segmentation.

Finally, we add both audio features and semantic-level cognitive consensus to the model, i.e. the proposed C3N. It can be noticed that with the cognitive consensus-based cross-modal alignment, the performance of the model improves remarkably under all four backbone settings. In particular, under the MS3 setting, the gains of mIoU are 3.33$\%$, 2.75$\%$, 3.38$\%$, 2.91$\%$ respectively and F-score increases by 0.038, 0.022, 0.034 and 0.026. For the S4 setting, mIoU and F-score metrics generally rise by around 1$\%$ and 0.01 in all four experimental groups, respectively. The above results demonstrate the audio-visual cognitive consensus inference and feedback facilitates explicit cross-modal alignment and improves AVS model performance.

\begin{figure*}[!t]
\centering
\includegraphics[width=0.95\linewidth]{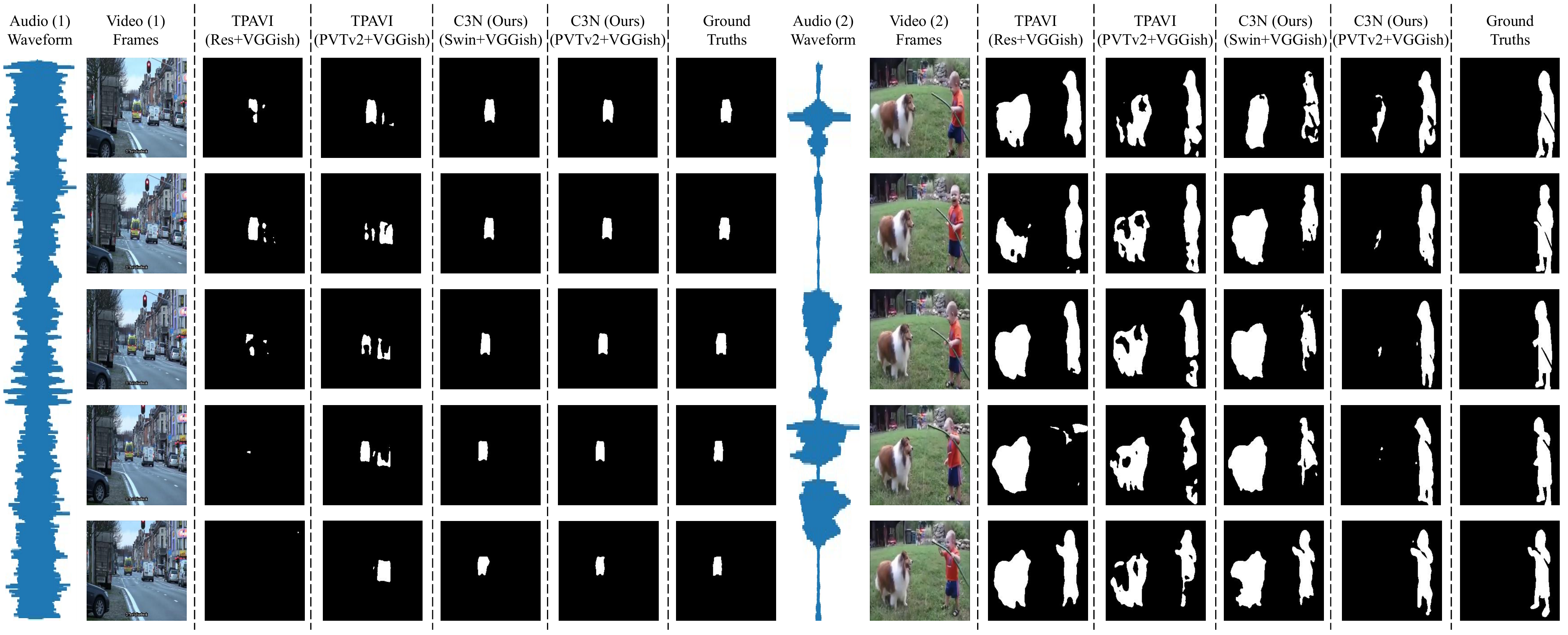}
\caption{Audio-Visual Segmentation examples of TPAVI \cite{zhou2022audio} and our C3N on the val/test set of AVSBench dataset.}
\label{fig:4}
\end{figure*}

\begin{table}[t]
\centering
\caption{Ablation experiments on hyperparameters $\alpha$ and $\beta$. The results are derived on the validation set of AVSBench under the Swin+PANNs backbone setting.}
\label{tab:3}
\setlength{\tabcolsep}{2.0mm}
{
\begin{tabular}{c|c|c|c|c|c}
\toprule
\multicolumn{6}{c}{Swin+PANNs}\\  \hline \hline
     \multicolumn{1}{c}{\multirow{2}{*}{$\alpha$}}     & 
     \multicolumn{1}{c|}{\multirow{2}{*}{$\beta$}}     &
     \multicolumn{2}{c|}{mIoU}                     & \multicolumn{2}{c}{F-score} \\ \cline{3-6}
  \multicolumn{1}{l}{}  & \multicolumn{1}{l|}{} & \multicolumn{1}{c|}{S4}              & \multicolumn{1}{c|}{MS3} & \multicolumn{1}{c|}{S4}              & \multicolumn{1}{c}{MS3}\\ \hline
\multicolumn{1}{c}{0.10}  &\multicolumn{1}{c|}{0.10}  & \textbf{81.36} & \textbf{61.10} & \textbf{0.894}  & 0.721 \\
\multicolumn{1}{c}{0.25}  &\multicolumn{1}{c|}{0.25}  &  81.08  &  60.74  & 0.892  & \textbf{0.723} \\
\multicolumn{1}{c}{0.10}  &\multicolumn{1}{c|}{0.25}  &  81.01  & 60.69  &  0.891 & 0.718 \\
\multicolumn{1}{c}{0.25}  &\multicolumn{1}{c|}{0.10}  &  80.73  &  60.47 &  0.890 &  0.719 \\ 
\bottomrule

\end{tabular}
}
\end{table}
\subsubsection{Analysis of Hyperparameters}

The two hyperparameters involved in the C3N framework are balance coefficients $\alpha$ and $\beta$ in Equation (\ref{eq:(4)}). Thus, we conduct multiple settings for the two parameters on the validation set of the S4 and MS3 scenarios under the Swin+PANNs backbone setting. The detailed quantitive results are shown in Table \ref{tab:3}. Experimental results show that the F-score metrics under different settings fluctuate within a relatively small range. Furthermore, the model performance is the overall best when $\alpha$=$\beta$=0.10. 


\subsubsection{Analysis of Model Complexity}

\begin{table}[!t]
\centering
\caption{Parameter and inference FLOPs of the baseline and C3N methods under the PVTv2+VGGish backbone combination.}
\label{tab:complexity}
\setlength{\tabcolsep}{4.0mm}
{
\begin{tabular}{l|c|c|c}
\toprule
 \textit{Complexity} & Baseline &  C3N &  \textbf{$\Delta$($\%$)} \\ 
 \hline\hline
 FLOPs (GB) & 161.36 &  161.39 &  \textbf{0.019} \\ \hline
\#Params (MB) & 171.46 & 177.15 &  \textbf{3.32} \\ 
\bottomrule
\end{tabular}
}
\end{table}

In Table \ref{tab:complexity}, we evaluate the model complexity of the baseline and C3N methods under the same backbone setting, where the baseline means removing the proposed C3IM and CCAM from the C3N model. $\Delta$ means the increased FLOPs/Params of our C3N as a proportion of the baseline model. The inference FLOPs and the number of parameters of the C3N increase by 0.03GB and 5.69 MB, which account for 0.019$\%$ and 3.32$\%$ of the baseline, respectively. Considering the improvement of the model performance, the increased model complexity is tolerable.

\subsection{In-depth Analysis}
\subsubsection{Qualitative analysis}
In Fig. \ref{fig:4}, we present two AVS examples on the validation and test set of the AVSBench dataset for a qualitative comparison between the baseline method TPAVI \cite{zhou2022audio} and our C3N. We show segmentation masks of two TPAVI models (i.e. \text{ResNet-50+VGGish} version and \text{PVTv2+VGGish} version) and two C3N models (i.e. \text{Swin+VGGish} version and \text{PVTv2+VGGish} version).

As shown in the left panel, there is a sounding ambulance with an alarm lamp and a normal van in the image. The Res+VGGish and PVTv2+VGGish version TPAVI models either fail to segment the sounding ambulance or incorrectly segment the normal van. The reason is that the dense feature-level interactions of the TPAVI model lead to the mislocalization of the representationally similar but semantically different object. However, our C3N utilizes cognitive consensus to align the audio and visual modalities from the semantic level and make correct localization and precise segmentation.

\begin{figure*}[!t]
\centering
\includegraphics[width=0.95\linewidth]{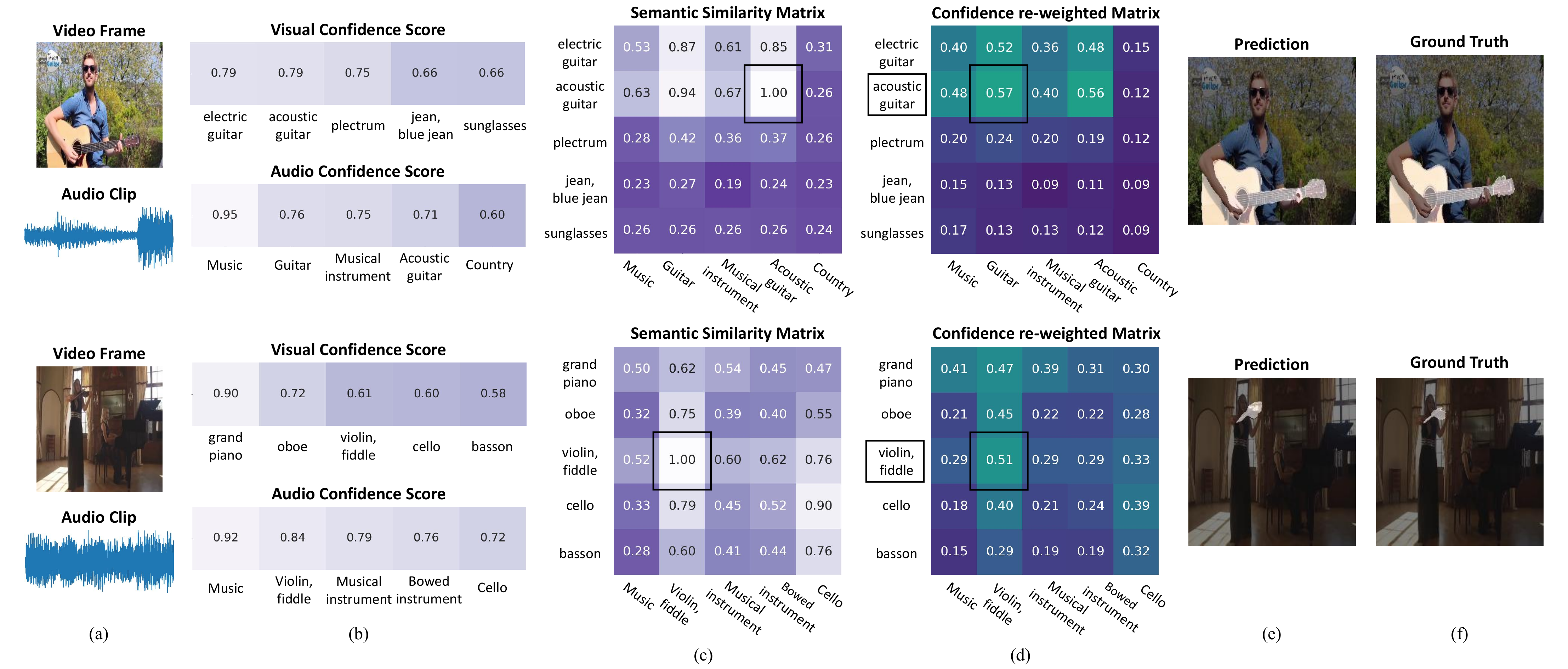}
\caption{Visualization of cross-modal cognitive consensus inference process (top-5 confidence audio/visual classes are shown).}
\label{fig:5}
\end{figure*}

For the right sample of Fig. \ref{fig:4}, the video shows a child playing with a dog as the child continuously makes giggling sounds and the dog is silent. The example is hard because the color of the dog is significantly different from the background, which may mislead the model into incorrectly segmenting the dog as well. The TPAVI model (Res+VGGish) even completely ignores the sounding child sometimes. The TPAVI (PVTv2+VGGish) and C3N (Swin+VGGish) segment the dog and child simultaneously. The reason why our C3N (Swin+VGGish) makes false segmentation of the dog is that the Swin generates unbalanced confidence scores and overly attends to the dog in the image, despite the dog and kid should be equally treated without audio guidance, which causes the unified-modal label shifts to the dog. However, the C3N (PVTv2+VGGish) overcomes distraction from the other object and accurately segments the sounding child.

\subsubsection{Effect of cross-modal cognitive consensus inference}

The C3IM integrates the audio and visual classification confidence scores and label embedding similarities to extract cognitive consensus. In Fig. \ref{fig:5}, we present two examples of the cross-modal cognitive consensus inference process. For ease of illustration, we only show top-5 confidence classes for audio and visual modalities. Note that the audio and visual confidence scores correspond to ${(c_{j}^{A})}^{\alpha }$ and ${(c_{k}^{V})}^{\beta }$ in Equation. (\ref{eq:(4)}). The highest semantic similarity value, confidence re-weighted value, and the corresponding unified-modal label are framed.

In the top example, the image shows a man playing the guitar, and the audio is the sound of the guitar. Column (b) shows that the visual classification head can not tell if it is an electric guitar or an acoustic guitar, and the man's clothing also has high confidence. In column (c), the audio-visual label similarities are calculated to establish initial semantic-level correlations. In column (d), the confidence re-weighted matrix is obtained by multiplying audio/visual confidence and cross-modal label similarities. It is noticeable that the visual class corresponding to the maximum value is \text{acoustic guitar}, which is consistent with the input audio and visual information. Based on the unified-modal label, the model perceives the object of interest. After that, columns (e) and (f) show the model accurately segments the sounding acoustic guitar.

In the bottom example, the sounding object is the violin. This example is more difficult because other prominent targets are present in the visual scene, such as humans and the piano. The sounding violin occupies only a small part of the frame. Column (b) presents the \text{grand piano} has the highest confidence and the model incorrectly regards the violin as an oboe. However, after combining it with the audio confidence and semantic similarities, the inferred cognitive consensus still emphasizes the correct \text{violin, fiddle} class, which is shown in column (d). According to such guidance information, the model correctly segments the violin. The above visualization results confirm the validity of our C3IM.

\subsubsection{Effect of cognitive consensus guided attention}

\begin{figure*}[t]
\centering
\includegraphics[width=0.95\linewidth]{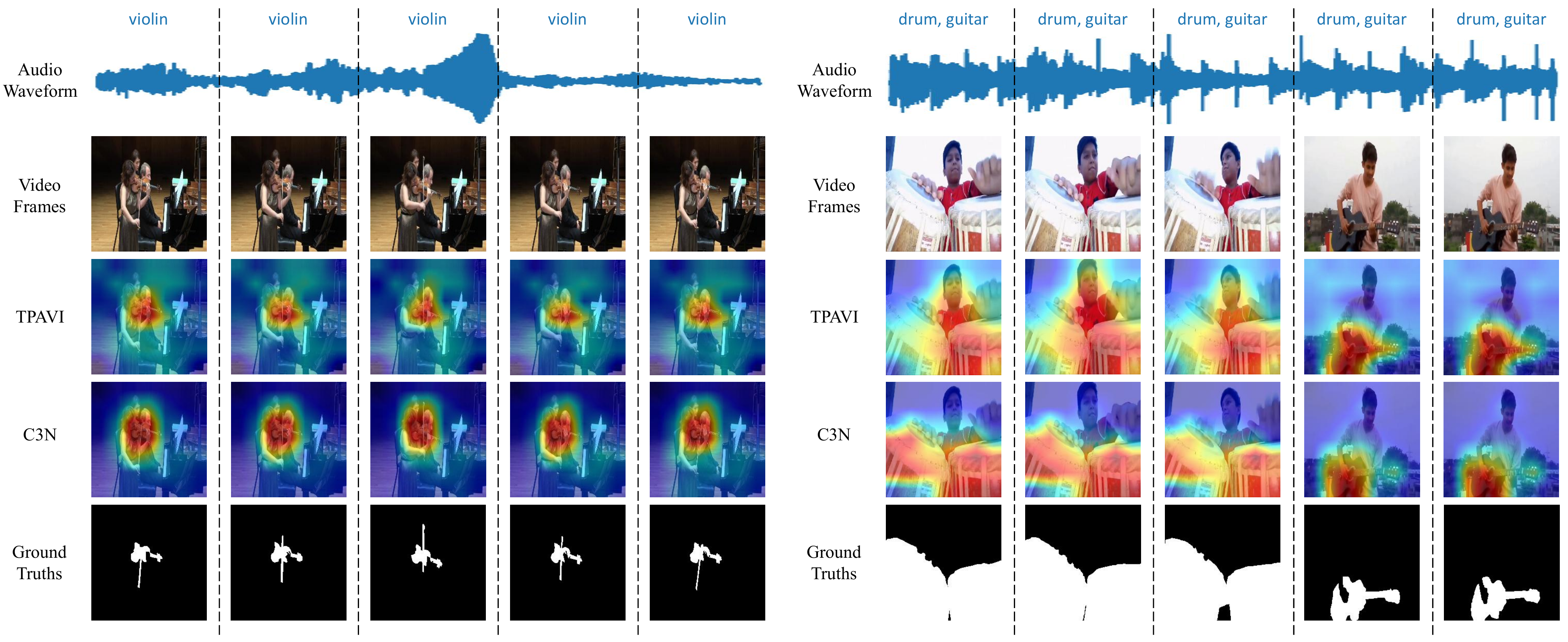}
\caption{Visualization of heat maps that come from the audio-visual attention module of TPAVI, and CCAM of our C3N.}
\label{fig:6}
\end{figure*}

The Cognitive Consensus guided Attention Module (CCAM) aims to highlight the local feature elements of the sounding object based on inferred cross-modal cognitive consensus. For comparison, we visualize the attention matrices within the audio-visual attention module of TPAVI \cite{zhou2022audio} and the spatial attention map within the CCAM of the proposed C3N. In practice, we remove the non-linear activations of CCAM.

In Fig. \ref{fig:6}, the heat maps of attention maps of the TPAVI baseline and the CCAM are shown in the third and fourth rows, respectively. Note that in the fourth row, only the semantic-level alignment of audio and visual modalities is performed, while the third row shows the audio-visual feature-level attention. In the left example, the violin produces sounds and other objects keep silent. In the third row, although the TPAVI baseline locates the violin, other irrelevant regions are also highlighted to varying degrees. Nevertheless, in the fourth row, only the sounding violin is emphasized in the spatial attention map of the CCAM. In the right sample, the situation is more complex, as two instrument sounds are mixed, and the scene switches in the video. It requires the model to accurately locate the sounding objects in distinctive scenes. The TPAVI baseline incorrectly highlights drummers. However, with the help of cognitive consensus, the CCAM heat maps locate the sounding drum and guitar in the first three frames and the last two frames respectively, which is consistent with ground truths. The above results demonstrate the CCAM allows accurately localizing the sounding object only relying on semantic-level cognitive consensus.

\subsubsection{Effect of cross-modal feature fusion}

\begin{figure}[!t]
\centering
\includegraphics[width=1.0\linewidth]{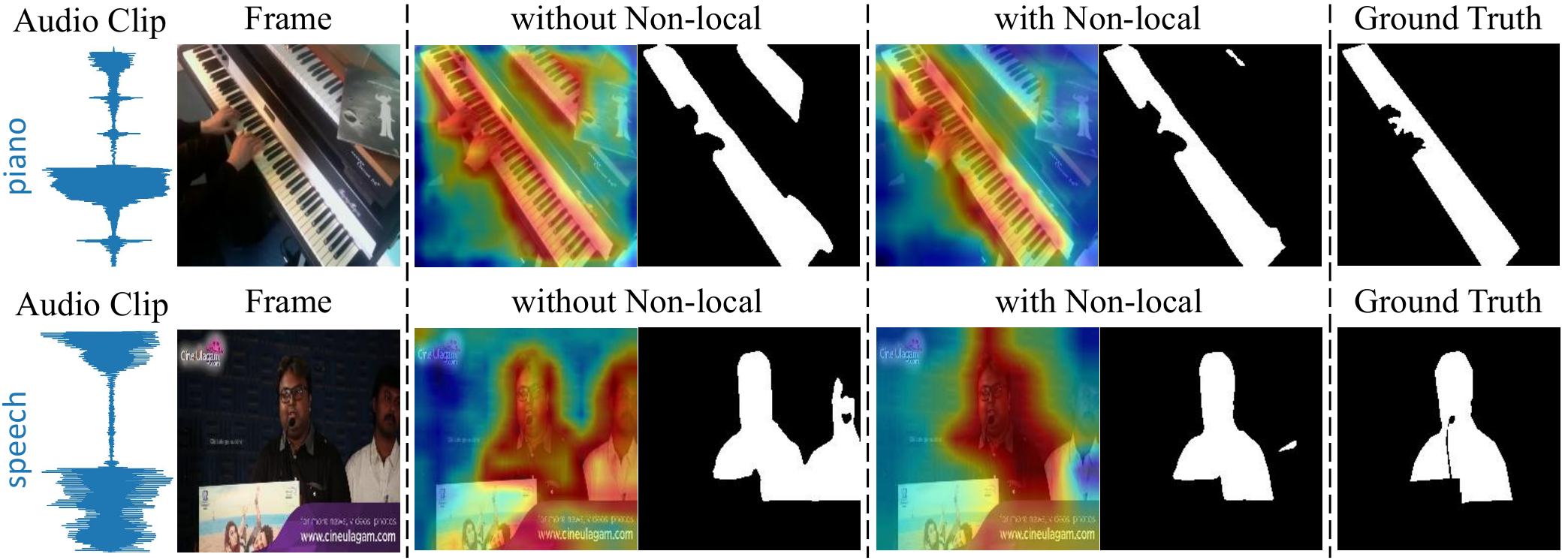}
\caption{Visualization of the feature heatmaps and segmentation predictions of our C3N model without or with the cross-modal Non-local block-based feature fusion.}
\label{fig:feature_level}
\end{figure}

The role of the cross-modal Non-local block-based feature fusion is to perform dense audio-visual feature-level interaction to calibrate the visual features via the discriminative audio features, which enforces highlighting the visual context of sounding objects to distinguish the sounding and silent objects with similar semantics. In Fig. \ref{fig:feature_level}, we show the visual feature heatmaps obtained by averaging the feature and the segmentation predictions output by our C3N with or without the cross-modal Non-local blocks. In the first row, there are two men with the same semantic label in a frame while one of them is speaking. The C3N without the cross-modal Non-local blocks incorrectly highlights and segments both of them. Our C3N with the cross-modal Non-local blocks identifies the speaking man by calibrating the visual features via the discriminative audio features to make correct segmentation. In the second row, there are two pianos, one of which is being played. The C3N without the cross-modal Non-local blocks incorrectly highlights and segments the silent piano. Whereas the C3N with the cross-modal Non-local blocks highlights the piano being played by hands and predicts the accurate segmentation mask.

\subsubsection{Failure cases}

\begin{figure}[t]
\centering
\includegraphics[width=0.90\linewidth]{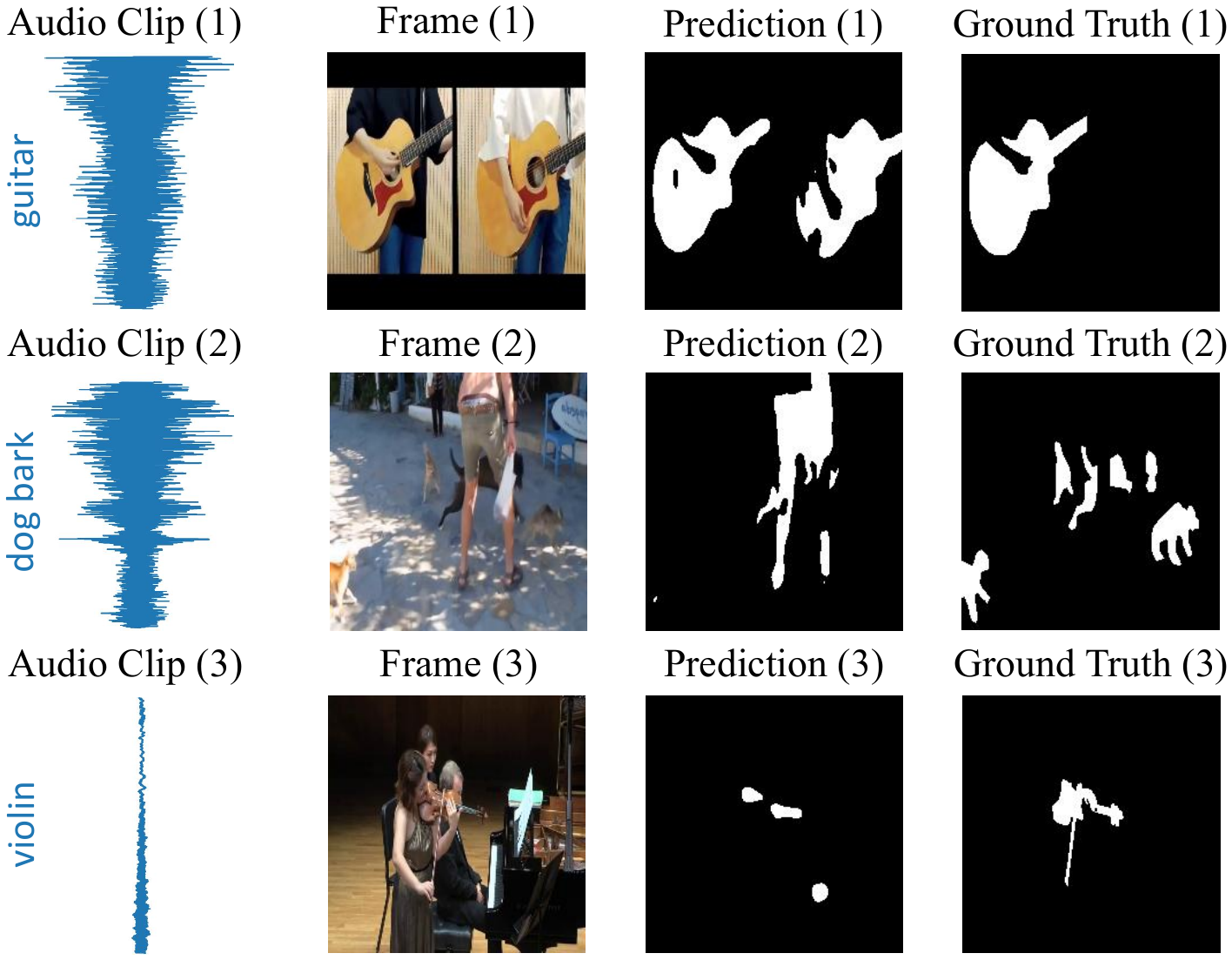}
\caption{Failure cases of our C3N under the MS3 setting of AVSBench dataset.}
\label{fig:7}
\end{figure}

We visualize three failure cases in Fig. \ref{fig:7}. In the first row, the failure example is caused by highly similar visual appearances with identical semantics. Specifically, the semantics and appearances of the two guitars are highly similar and both have a hand on them, and it is difficult for the model to tell which guitar is sounding. In the second row, the failure example is caused by object occlusion. Specifically, the barking dog is covered by the man in a large area, which causes the C3N model to mislocate and output failure prediction. In the third row, the volume of the input audio is so low that the audio features lose the discriminability towards the sounding object, so the model cannot localize the sounding violin and just outputs the mask with several tiny positive areas. In the above cases, it is difficult to identify and segment the sounding object based on the limited information of a single frame. In the future, we will concentrate on mining the inter-frame temporal context compensational information of the audio-visual inputs to address problems in these cases.

\subsubsection{Evaluation under different genres of videos}

\begin{table*}[ht]
\centering
\caption{The per-category mIoU and F-score performance of our C3N (PVTv2+PANNs) under the S4 setting.}
\label{tab:category}
\scalebox{0.81}{
\begin{tabular}{p{0.9cm}<{\centering}|p{0.9cm}<{\centering}|p{0.9cm}<{\centering}|p{0.9cm}<{\centering}|p{0.9cm}<{\centering}|p{0.9cm}<{\centering}|p{0.9cm}<{\centering}|p{0.9cm}<{\centering}|p{0.9cm}<{\centering}|p{0.9cm}<{\centering}|p{0.9cm}<{\centering}|p{0.9cm}<{\centering}|p{0.9cm}<{\centering}|p{0.9cm}<{\centering}|p{1.4cm}<{\centering}|p{1.4cm}<{\centering}}
\toprule
\multicolumn{2}{c|}{ambulance siren} & \multicolumn{2}{c|}{baby laughter}  & \multicolumn{2}{c|}{gun shooting} & \multicolumn{2}{c|}{cat meowing}    & \multicolumn{2}{c|}{chainsawing trees} & \multicolumn{2}{c|}{coyote howling} & \multicolumn{2}{c|}{dog barking}    & \multicolumn{2}{c}{driving buses}  \\ \hline
mIoU  & F-score & mIoU  & F-score & mIoU   & F-score  & mIoU & F-score & mIoU    & F-score  & mIoU  & F-score & mIoU  & F-score & mIoU  & F-score \\ \hline
79.81  & 0.863   & 82.62 & 0.894   & 72.57  & 0.851    & 87.77 & 0.934   & 65.45  & 0.789    & 83.36 & 0.914   & 87.26 & 0.927   & 84.08 & 0.891  \\ \hline
\hline
\multicolumn{2}{c|}{female singing} & \multicolumn{2}{c|}{helicopter}  & \multicolumn{2}{c|}{horse clip-clop} & \multicolumn{2}{c|}{lawn mowing}    & \multicolumn{2}{c|}{lions roaring} & \multicolumn{2}{c|}{male speech} & \multicolumn{2}{c|}{bird singing}    & \multicolumn{2}{c}{ playing guitar}  \\ \hline
mIoU  & F-score & mIoU  & F-score & mIoU   & F-score  & mIoU & F-score & mIoU    & F-score  & mIoU  & F-score & mIoU  & F-score & mIoU  & F-score  \\ \hline
81.70  & 0.878   & 76.96 & 0.880   & 81.16   & 0.895   & 86.72 & 0.931  & 91.13   & 0.959    & 90.61 & 0.958   & 87.96 & 0.949   & 87.71 & 0.946   \\ \hline
\hline
\multicolumn{2}{c|}{playing glockenspiel} & \multicolumn{2}{c|}{playing piano}  & \multicolumn{2}{c|}{playing tabla} & \multicolumn{2}{c|}{playing ukulele}    & \multicolumn{2}{c|}{playing violin} & \multicolumn{2}{c|}{race car} & \multicolumn{2}{c|}{typing keyboard}    & \multicolumn{2}{c}{\textbf{Class Average}}  \\ \hline
mIoU  & F-score & mIoU  & F-score & mIoU   & F-score  & mIoU & F-score & mIoU    & F-score  & mIoU  & F-score & mIoU  & F-score & mIoU  & F-score  \\ \hline
83.45  & 0.934   & 87.30 & 0.940   & 87.07   & 0.920    & 76.59 & 0.873  & 74.35   & 0.861    & 85.74 & 0.924   & 88.74 & 0.949   & \textbf{83.05$\pm$6.21} & \textbf{0.907$\pm$0.041}  \\ \bottomrule
\end{tabular}}
\end{table*}

To evaluate the effectiveness of our C3N under different genres of videos as inputs, we conduct the per-category performance test on 23 categories for our C3N (PVTv2+PANNs) under the S4 setting, which has the video category annotations, and the average and standard deviation of mIoU and F-score metrics across all categories are also reported. As shown in Table \ref{tab:category}, our C3N (PVTv2+PANNs) achieves an average mIoU of 83.05$\%$ and an average F-score of 0.907 across all 23 categories, with their respective standard deviation of 6.21$\%$ and 0.04, which demonstrates that the performance of our C3N fluctuates within a relatively small range with video category changes.

\subsubsection{Evaluation under varying sound qualities}
\begin{table}[t]
\centering
\caption{The mIoU and F-score performance of our C3N (PVTv2+PANNs) with the incorporation of varying degrees of Gaussian Noise to the audio under the S4 and MS3 settings. ``SNR" represents the signal-to-noise ratio.}
\label{tab:noise}
\begin{tabular}{p{1.5cm}<{\centering}|p{1.0cm}<{\centering}|p{1.0cm}<{\centering}|p{1.0cm}<{\centering}|p{1.0cm}<{\centering}}
\toprule
\multirow{2}{*}{SNR (dB)} & \multicolumn{2}{c|}{mIoU}     & \multicolumn{2}{c}{F-score}  \\ \cline{2-5} 
                     & S4 & MS3 & S4 & MS3 \\ \hline\hline
No noise             &  \textbf{82.94}  & \textbf{61.67}   &  \textbf{0.906}  &  \textbf{0.713}   \\ \hline
20dB                 &  82.86  &  61.52   &  0.905  &   0.713  \\ \hline
10dB                 &  82.79  &  60.60  &  0.904  &   0.703  \\ \hline
0dB                  &  82.65  &   59.16  & 0.903   &   0.693  \\ \hline
-10dB                &  82.27  &  55.23   &  0.900 &   0.674  \\ \hline
-20dB                &  81.50  &  51.08  &   0.896  &   0.645  \\ \bottomrule
\end{tabular}
\end{table}

To evaluate the adaptivity of our method under varying sound qualities, we add Gaussian white noise of a variety of signal-to-noise ratios (SNR) to the input audio to simulate different quality audio signals. As demonstrated in Table \ref{tab:noise}, the segmentation performance of our C3N declines with the SNR decreasing. Under the S4 setting, since there is only one sounding object in the video, our C3N is slightly influenced by the added noise, i.e. the mIoU and F-score decrease in the range of 1.5$\%$ mIoU and 0.01 F-score with 20dB to -20dB SNR audio inputs, respectively. Due to the presence of multiple sounding objects in a video under the MS3 setting, our C3N is more sensitive to audio quality compared to the S4 setting. When the SNR decreases from 20dB to 0dB, the performance of our C3N drops less than 3$\%$ mIoU and 0.02 F-score compared with C3N with clear audio inputs. Although the performance of C3N significantly degrades when the added noise power is greater than the clear audio (i.e. SNR$<$0dB), our C3N still achieves 55.23$\%$ mIoU and 0.674 F-score with -10dB SNR audio, and 51.08$\%$ mIoU and 0.645 F-score with -20dB SNR audio. The above results demonstrate the robustness of the proposed C3N under varying sound qualities.

\subsubsection{Evaluation under varying visual qualities}

\begin{figure}[t]
\centering
\includegraphics[width=1.0\linewidth]{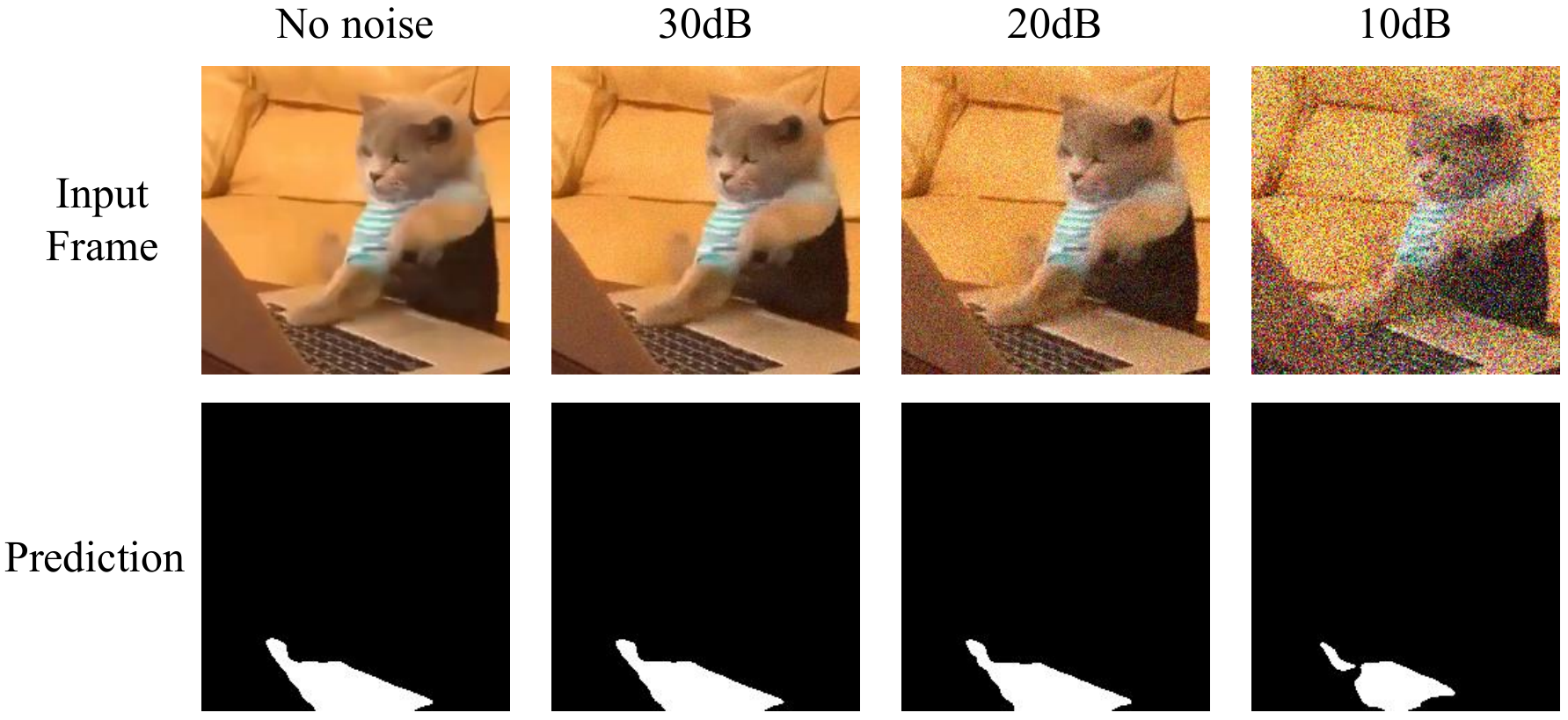}
\caption{Visualizations of the input video frame with Gaussian Noise of different signal-to-noise ratios (SNR), and their corresponding segmentation prediction output by C3N (PVTv2+PANNs).}
\label{fig:image_noise}
\end{figure}

\begin{table}[t]
\centering
\caption{The mIoU and F-score of our C3N (PVTv2+PANNs) with the incorporation of varying degrees of Gaussian Noise to the video frames under the S4 and MS3 settings. ``SNR" represents the signal-to-noise ratio.}
\label{tab:noise_image}
\begin{tabular}{p{1.5cm}<{\centering}|p{1.0cm}<{\centering}|p{1.0cm}<{\centering}|p{1.0cm}<{\centering}|p{1.0cm}<{\centering}}
\toprule
\multirow{2}{*}{SNR (dB)} & \multicolumn{2}{c|}{mIoU}     & \multicolumn{2}{c}{F-score}  \\ \cline{2-5} 
                     & S4 & MS3 & S4 & MS3 \\ \hline\hline
No noise             &  \textbf{82.94}  & \textbf{61.67}   &  \textbf{0.906}  &  \textbf{0.713}   \\ \hline
30dB                 &  82.58  &  61.59   & 0.904  &  0.712   \\ \hline
20dB                 &  80.81  &  59.40  &  0.894  &   0.697  \\ \hline
10dB                  &  74.06  &  52.09  &  0.849 &   0.629 \\  \bottomrule
\end{tabular}
\end{table}

To evaluate the adaptivity of our method under noisy visual environments, we add Gaussian noise of varying signal-to-noise ratios (SNR) to the input video frames to simulate different quality videos. In Fig. \ref{fig:image_noise}, we show video frames with the noise of different signal-to-noise ratios (SNR) and their respective segmentation prediction output by our C3N (PVTv2+PANNs). When the SNR of the input frame is 30dB or 20dB, our C3N outputs accurate segmentation masks. However, when SNR is decreased to 10dB, the frame is severely distorted, and the quality of the predicted mask is significantly lower compared to the clear frame input. As shown in Table \ref{tab:noise_image}, the C3N performance declines with the SNR decreasing. When the SNR drops to 20 dB, the mIoU and F-score slightly reduce by 2.13 $\%$ and 0.012 under the S4 setting and 2.27$\%$ and 0.016 under the MS3 setting. Despite the performance significantly drops when SNR is decreased to 10dB, our C3N still achieves 74.06$\%$ mIoU and 0.849 F-score under the S4 setting, and 52.09$\%$ mIoU and 0.629 F-score under the MS3 setting. The above experimental results demonstrate the noise robustness of our C3N when encountering not severe visual noise.

\subsubsection{Comparison between C3IM and other cross-modal interaction modules}

\begin{table}[t]
\centering
\caption{The mIoU and F-score results of our C3N (PVTv2+PANNs) using the proposed C3IM or other alternative cross-modal interaction modules under the S4 and MS3 settings.}
\label{tab:Comparison between C3IM and other modules}
\begin{tabular}{p{2.2cm}<{\centering}|p{1.0cm}<{\centering}|p{1.0cm}<{\centering}|p{1.0cm}<{\centering}|p{1.0cm}<{\centering}}
\toprule
\multirow{2}{*}{Modules} & \multicolumn{2}{c|}{mIoU}     & \multicolumn{2}{c}{F-score}  \\ \cline{2-5} 
                     & S4 & MS3 & S4 & MS3 \\ \hline\hline
C3IM           &  \textbf{82.94}  & \textbf{61.67}   &  \textbf{0.906}  &  \textbf{0.713}   \\ \hline
MHA \cite{chen2024unraveling,gao2024avsegformer}           &  82.16  &  61.20  & 0.901  & 0.706   \\ \hline
BFM  \cite{yang2024cooperation}           &  81.46  &  59.50  &  0.899 & 0.701  \\ \hline
TPAVI \cite{zhou2022audio}                  &  80.81 & 58.69  & 0.903  & 0.684 \\  \bottomrule
\end{tabular}
\end{table}

To further demonstrate the superiority of the proposed C3IM, we conduct detailed comparative experiments that replace the C3IM in our C3N (PVTv2+PANNs) with other cross-model interaction modules. In detail, we choose three recent cross-modal multi-head attention (MHA) module used in \cite{chen2024unraveling,gao2024avsegformer}, Bilateral-Fusion Module (BFM) used in \cite{yang2024cooperation}, and the temporal pixel-wise audio-visual interaction (TPAVI) module used in \cite{zhou2022audio} as alternatives to the C3IM. Experimental results in Table \ref{tab:Comparison between C3IM and other modules} show our C3IM outperforms other powerful cross-model interaction modules under the S4 and MS3 settings. Specifically, our C3IM achieves higher mIoU than the MHA, BFM, and TPAVI by 0.78$\%$, 1.48$\%$, and 2.13$\%$, respectively. Under the MS3 setting, the C3IM surpasses the MHA, BFM, and TPAVI by 0.47$\%$, 2.17$\%$, and 2.98$\%$ in mIoU, and 0.007, 0.012, and 0.029 in F-score, respectively.

\section{Conclusion}
In this paper, we have proposed Cross-modal Cognitive Consensus guided Network (C3N), a novel framework for Audio-Visual Segmentation that exploits semantic-level information for explicit cross-modal alignment. We obtain the unified-modal label by integrating audio/visual classification confidence and semantic similarities via a Cross-modal Cognitive Consensus Inference Module (C3IM). In addition, we develop a Cognitive Consensus guided Attention Module (CCAM) to highlight the local features corresponding to the object of interest depending on the global cognitive consensus guidance. Extensive experiments verify the effectiveness of our method and its superiority over state-of-the-art methods.

For the AVS task, the sounding objects in each frame of the video are frequently temporally correlated. Thus, it is also necessary to design inter-frame interaction methods to enhance the frame-to-frame correlations of the segmented objects. The design of the inter-frame interaction method needs to focus on two aspects: On the one hand, when the appearance of the sounding object suddenly changes such as it is highly occluded, the inter-frame compensational information should be utilized to accurately identify and segment the sounding object. On the other hand, when the audio abruptly changes, the model needs to capture the inter-frame discrepancy information to exclude the irrelevant objects in the video. In the future, we will further explore the modeling of the frame-to-frame correlations of the sounding objects. 

\section*{Acknowledgments}
This work was supported in part by the National Science and Technology Major Project (2021ZD0112001), the National Natural Science Foundation of China (No. U23A20286), the Independent Research Project of Civil Aviation Flight Technology and Flight Safety Key Laboratory (FZ2022ZZ06), and the Natural Science Foundation of Sichuan Province (2023NSFSC1972).

\bibliographystyle{IEEEtran}
\bibliography{references}{}

\begin{IEEEbiography}[{\includegraphics[width=1in,height=1.25in,clip,keepaspectratio]{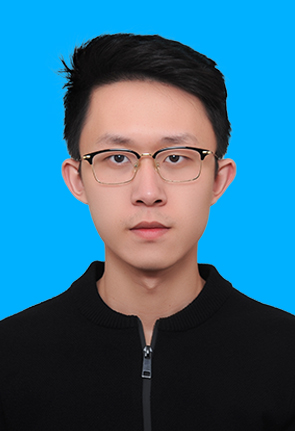}}]
{Zhaofeng Shi} received the B.E. degree in Electronic Information Engineering at the University of Electronic Science and Technology of China (UESTC) in 2021 and completed his master's studies in 2023. Now he is pursuing his Ph.D. degree in Information and Communication Engineering. His main research interests include egocentric understanding, multi-modal processing, and computer vision.
\end{IEEEbiography}

\begin{IEEEbiography}[{\includegraphics[width=1in,height=1.25in,clip,keepaspectratio]{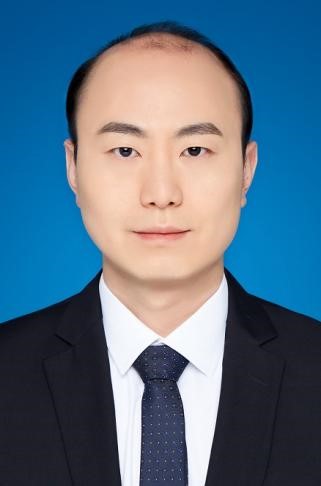}}]
{Qingbo Wu} (Member, IEEE) received the Ph.D. degree in signal and information processing from the University of Electronic Science and Technology of China in 2015. From February 2014 to May 2014, he was a Research Assistant with the Image and Video Processing (IVP) Laboratory, Chinese University of Hong Kong. From October 2014 to October 2015, he served as a Visiting Scholar with the Image and Vision Computing (IVC) Laboratory, University of Waterloo. He is currently an Associate Professor with the School of Information and Communication Engineering, University of Electronic Science and Technology of China. His research interests include image/video coding, quality evaluation, perceptual modeling and processing. He has served as Area Chair for ACM MM 2024, VCIP 2016, Session Chair for ACM MM 2021, ICMCT 2022, TPC/PC member of AAAI 2021-2023, APSIPA ASC 2020-2021, CICAI 2021-2023. He was also a Guest Editor of Remote Sensing and Frontiers in Neuroscience.
\end{IEEEbiography}

\begin{IEEEbiography}[{\includegraphics[width=1in,height=1.25in,clip,keepaspectratio]{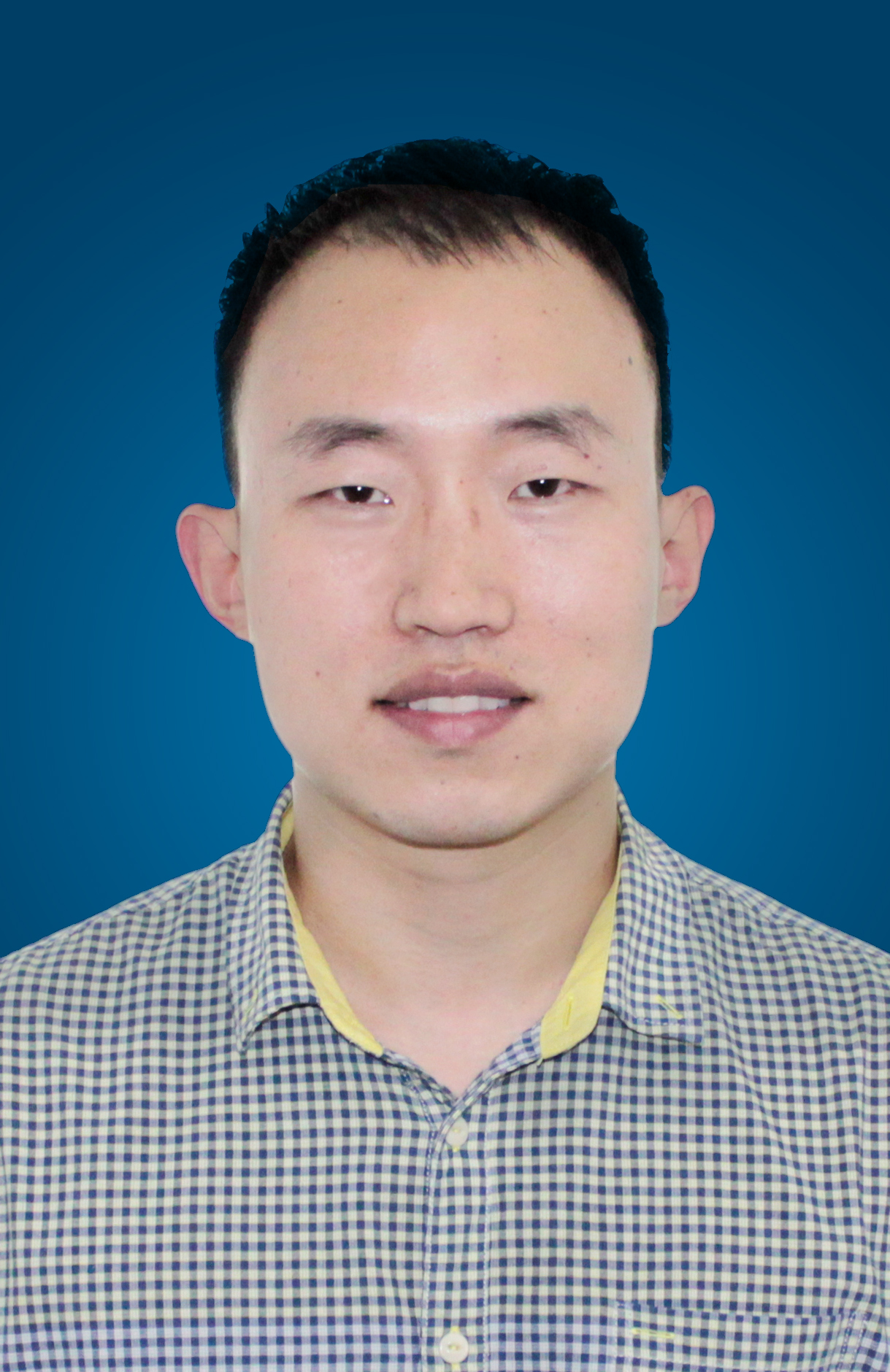}}]
{Fanman Meng} (S’12–M’14) received the Ph.D. degree in signal and information processing from the University of Electronic Science and Technology of China, Chengdu, China, in 2014. From 2013 to 2014, he was a Research Assistant with the Division of Visual and Interactive Computing, Nanyang Technological University, Singapore. He is currently Professor with the School of Information and Communication Engineering, University of Electronic Science and Technology of China. He has authored or co-authored numerous technical articles in well-known international journals and conferences. His current research interests include image segmentation and object detection. 

Dr. Meng is a member of the IEEE Circuits and Systems Society. He was a recipient of the Best Student Paper Honorable Mention Award at the 12th Asian Conference on Computer Vision, Singapore, in 2014, and the Top 10$\%$ Paper Award at the IEEE International Conference on Image Processing, Paris, France, in 2014.
\end{IEEEbiography}

\begin{IEEEbiography}[{\includegraphics[width=1in,height=1.25in,clip,keepaspectratio]{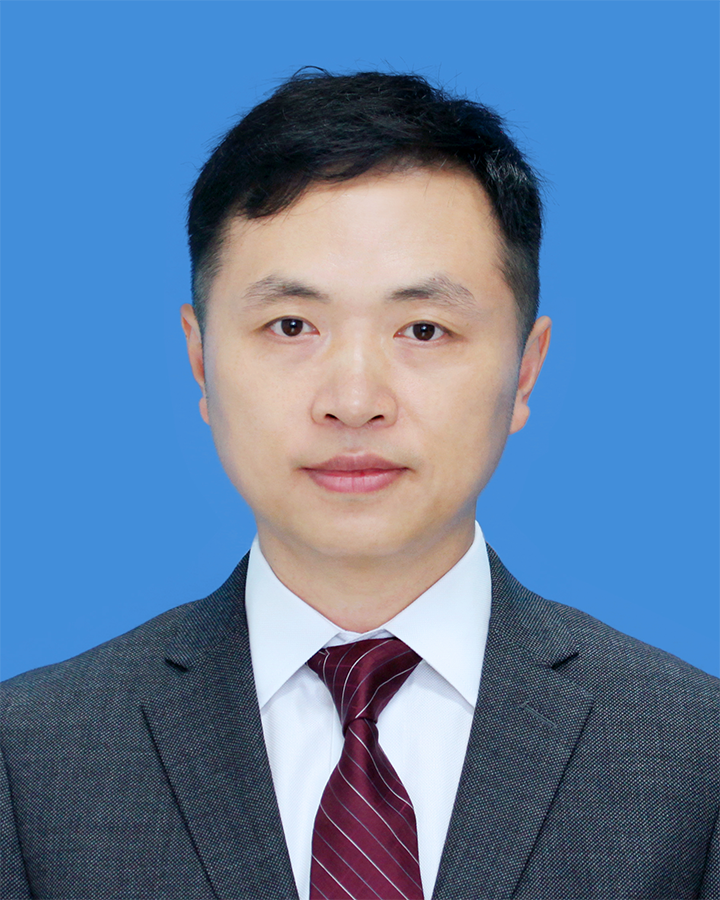}}]
{Linfeng Xu} (Member, IEEE) received the Ph.D. degree in Signal and Information Processing from the School of Electronic Engineering, University of Electronic Science and Technology of China (UESTC), Chengdu, China, in 2014. From December 2014 to December 2015, he was with the Ubiquitous Multimedia Laboratory, the State University of New York at Buffalo, USA, as a visiting scholar. He is currently an Associate Professor with the School of Information and Communication Engineering, UESTC. His research interests include machine learning, computer vision, visual signal processing, artificial intelligence theory and applications. He served as a Local Arrangement Chair for ISPACS 2010 and VCIP 2016.
\end{IEEEbiography}

\begin{IEEEbiography}[{\includegraphics[width=1in,height=1.25in,clip,keepaspectratio]{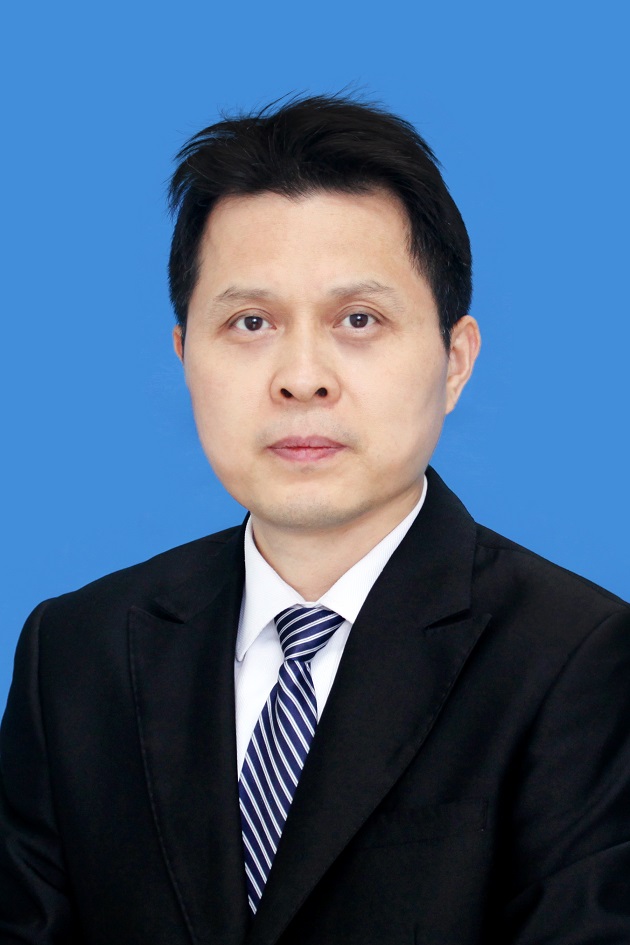}}]
{Hongliang Li} (SM’12) received his Ph.D. degree in Electronics and Information Engineering from Xi’an Jiaotong University, China, in 2005. From 2005 to 2006, he joined the visual signal processing and communication laboratory (VSPC) of the Chinese University of Hong Kong (CUHK) as a Research Associate. From 2006 to 2008, he was a Postdoctoral Fellow at the same laboratory in CUHK. He is currently a Professor in the School of Information and Communication Engineering, University of Electronic Science and Technology of China. His research interests include image and video processing, visual attention, object detection and segmentation, object recognition and parsing, multimedia content analysis, deep learning.

Dr. Li has authored or co-authored numerous technical articles in well-known international journals and conferences. He is a co-editor of a Springer book titled ``Video segmentation and its applications”. Dr. Li is involved in many professional activities. He received the 2019 and 2020 Best Associate Editor Awards for IEEE Transactions on Circuits and Systems for Video Technology (TCSVT), and the 2021 Best Editor Award for Journal on Visual Communication and Image Representation. He served as a Technical Program Chair for VCIP 2016 and PCM 2017, General Chairs for ISPACS 2017 and ISPACS 2010, a Publicity Chair for IEEE VCIP 2013, a Local Chair for the IEEE ICME 2014, Area Chairs for VCIP 2022 and 2021, and a Reviewer committee member for IEEE ISCAS from 2018 to 2022. He served as an Associate Editor of IEEE Transactions on Circuits and Systems for Video Technology (2018-2021). He is now an Associate Editor of Journal on Visual Communication and Image Representation, IEEE Open Journal of Circuits and Systems, and an Area Editor of Signal Processing: Image Communication (Elsevier Science). He is selected as the IEEE Circuits and Systems Society Distinguished Lecturer for 2022-2023.
\end{IEEEbiography}

\end{document}